\pdfoutput=1
\documentclass[useAMS,usenatbib]{mn2e}
\usepackage[T1]{fontenc}
\usepackage[latin1]{inputenc}
\usepackage{amsmath}
\usepackage{graphicx}
\usepackage{amssymb}
\usepackage{epsfig}
\usepackage[english]{babel}
\usepackage{bbm}
\title[Closure tests for mean field magnetohydrodynamics]{Closure tests for mean field magnetohydrodynamics using a self consistent
reduced model}
\author[V.V.Pipin and M.R.E.Proctor] {V.V.Pipin$^{1,2}$ and M.R.E. Proctor$^{2}$\thanks{E-mail:
pip@iszf.irk.ru (VVP); mrep@cam.ac.uk (MREP)}\\
$^{1}$Institute for Solar-Terrestrial Physics,
Siberian Division of Russian Academy of Sciences, 664033 Irkutsk,
Russia\\
$^{2}$Centre for Mathematical Sciences, University of Cambridge, Wilberforce Road, Cambridge CB3 0WA, UK}

\begin{document}

\date{Accepted \today. Received \today; in original form \today}

\pagerange{\pageref{firstpage}--\pageref{lastpage}} \pubyear{2008}

\maketitle

\label{firstpage}

\begin{abstract}
The mean electromotive force and $\alpha$ effect are computed for
a forced turbulent flow using a simple nonlinear dynamical model.
The results are used to check the applicability of two basic analytic
\textit{ansätze} of mean-field magnetohydrodynamics - the second order
correlation approximation (SOCA) and the $\tau$ approximation. In
the numerical simulations the effective Reynolds number $Re$ is $2-20$,
while the magnetic Prandtl number $P_{m}$ varies from $0.1$ to $10^{7}$.
We present evidence that the $\tau$ approximation may be appropriate
in dynamical regimes where there is a small-scale dynamo.  Catastrophic
quenching of the $\alpha$ effect is found for high $P_{m}$. Our
results indicate that for high $P_{m}$ SOCA gives a very large value
of the $\alpha$ coefficient compared with the {}``exact'' solution.
The discrepancy depends on the properties of the random force that
drives the flow, with a larger difference occuring for $\delta$-correlated
force compared with that for a steady random force.
\end{abstract}

\begin{keywords}
Dynamo theory, solar magnetic fields
\end{keywords}

\section{Introduction}

\label{sec:int} It is widely believed that magnetic field generation
in cosmic bodies is governed by turbulent motions of electrically
conducting fluids \citep{moff:78,park,wiess94,bra-sub:04}. One of
the most important outstanding problems of astrophysical magnetohydrodynamics
is to explain the phenomenon of large-scale magnetic activity which
is observed in a wide range of astrophysical objects, e.g. the Sun
and late-type stars, galaxies, accretion disks, etc. In these cases
the spatial and temporal scales of the generated magnetic fields can
greatly exceed those of the turbulent fluctuating velocity and magnetic
fields. According to mean-field magnetohydrodynamics \citep{moff:78,park,krarad80}
the evolution of the large-scale magnetic field $\overline{\mathbf{B}}$
in turbulent highly-conducting fluid with mean velocity $\overline{\mathbf{U}}$
is governed by \begin{equation}
\frac{\partial\overline{\mathbf{B}}}{\partial t}=\nabla\times\boldsymbol{\mathcal{E}}+\nabla\times(\overline{\mathbf{U}}\times\overline{\mathbf{B}})+\eta\nabla^{2}\overline{\mathbf{B}},\label{eq:1i}\end{equation}
 where the mean electromotive force, $\boldsymbol{\mathcal{E}}=\left\langle \mathbf{u}\times\mathbf{b}\right\rangle $
is given by the correlation between the fluctuating components of
the velocity field of the plasma, $\mathbf{u}$, and the fluctuating magnetic
fields, $\mathbf{b}$. We can expect a linear relationship between
the mean electromotive force and the local large-scale magnetic field,
if the assumption of scale-separation holds \citep{moff:78,Proct03}:
\begin{equation}
\boldsymbol{\mathcal{E}}_{i}=(\nabla\times\left\langle \mathbf{u}\times\mathbf{b}\right\rangle )_{i}={\alpha_{ij}}\overline{{B}}_{j}+{\beta}_{ijk}\frac{\partial\overline{B}_{i}}{\partial x_{j}}+...,\label{eq:2i}\end{equation}
 where $\boldsymbol{\alpha}$ and $\boldsymbol{\beta}$ are tensors
which are usually evaluated by considering the dynamic equations for
the small-scale velocity and magnetic fields. If we suppose that $\overline{\mathbf{U}}=0$,
these equations are \begin{eqnarray}
\frac{\partial\mathbf{b}}{\partial t} & = & \mathbf{\nabla}\left(\mathbf{u\times b}-\left\langle \mathbf{u}\times\mathbf{b}\right\rangle +\mathbf{u\times\overline{B}}\right)+\eta\mathbf{\nabla^{2}b},\label{eq:1a}\\
\frac{\partial\mathbf{u}}{\partial t} & = & \nu\mathbf{\nabla^{2}u}-\mathbf{\nabla}\left(p+\frac{\mathbf{b^{2}}}{2\mu}+\frac{\left(\mathbf{b\cdot}\overline{\mathbf{B}}\right)}{\mu}\right)\label{eq:1b}\\
 &  & +\nabla_{i}\left(\frac{1}{\mu}\mathbf{b}b^{i}-\mathbf{u}u^{i}\right)+\frac{1}{\mu}\left(\overline{\mathbf{B}}\cdot\mathbf{\nabla}\right)\mathbf{b}+\mathbf{f},\nonumber \end{eqnarray}
 where $p$ is the fluctuating pressure, $\mathbf{f}$ is the random
force driving the turbulence and $\eta,\nu$ are the molecular diffusivity
and viscosity, respectively.

We could also self-consistently include the effects of rotation, since
the Coriolis force is linear; this enhancement is left for a future
paper. 

\begin{center}
\begin{figure}

\begin{centering}
\includegraphics[width=0.7\linewidth]{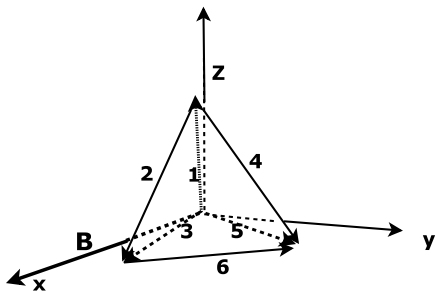} 
\par\end{centering}

\caption{\label{fig1} The geometry of the model.}

\end{figure}

\par\end{center}

It is known that the symmetric part of $\boldsymbol{\alpha}$ and
antisymmetric part of $\boldsymbol{\beta}$ in (\ref{eq:2i}) give
the source and diffusion terms of the mean magnetic field in (\ref{eq:1i}),
respectively.  The antisymmetric part of $\boldsymbol{\alpha}$
is usually interpreted as the mean pumping velocity and the symmetric
part of  $\boldsymbol{\beta}$ may contain the additional source term
$\overline{\mathbf{B}}$ (e.g. Rädler's, $\boldsymbol{\Omega}\times\mathbf{J}$
effect, \citep{rad69}). For the solar dynamo the symmetric part of
$\boldsymbol{\alpha}$ (or simply $\alpha$-effect) is a key ingredient
of most mean-fields models which claim to explain the large-scale
magnetic activity of the Sun.

There are currently two basic analytic methods for the approximate evaluation of $\boldsymbol{\mathcal{E}}$
and tensor coefficients in (\ref{eq:2i}) on the basis of  (\ref{eq:1a},\ref{eq:1b}).
The most usual method is the \emph{second order correlation approximation}
(SOCA)\citep{krarad80} which is also known as \textit{first order
smoothing} (FOSA) \citep{moff:78}. In this approximation,
all the nonlinear contributions of the fluctuating velocity and fluctuating
magnetic fields in (\ref{eq:1a},\ref{eq:1b}) are neglected. This
approximation has well-known limits to its accurate application. It
is good either for poorly conducting plasma (low $R_{m}$) or for
the weak turbulence case (low Strouhal number). Neither limit is
very appropriate in astrophysics where we have highly-conducting strongly
turbulent fluid. On the other hand the $\tau$-approximation, which uses
a higher order momentum closure and could be relevant for exploring
many common astrophysical situations, has no well defined mathematically
formulated limits. The particular variant of the $\tau$-approximation that is used in the paper will be described below.

In the paper by \cite{CHT06} the authors attempted to evaluate some
components of $\boldsymbol{\mathcal{E}}$ numerically. Their results
indicate a nontrivial dependence of the $\alpha$ effect on the basic
parameters of the turbulent flow, such as the correlation time, magnetic
Reynolds number and the helicity of the flow. Here we develop a kind
of shell model to explore some properties of mean-electromotive force
and especially the $\alpha$ effect in a wide range of turbulent regimes.
The model is useful for checking the basic approximations of  mean-field
magnetohydrodynamics as well, since it is simple enough to allow the
rapid calculation of different cases over a wide parameter range while
maintaining many properties of the full problem.

\begin{figure}

\begin{centering}
\includegraphics[width=.6\linewidth,angle=-90]{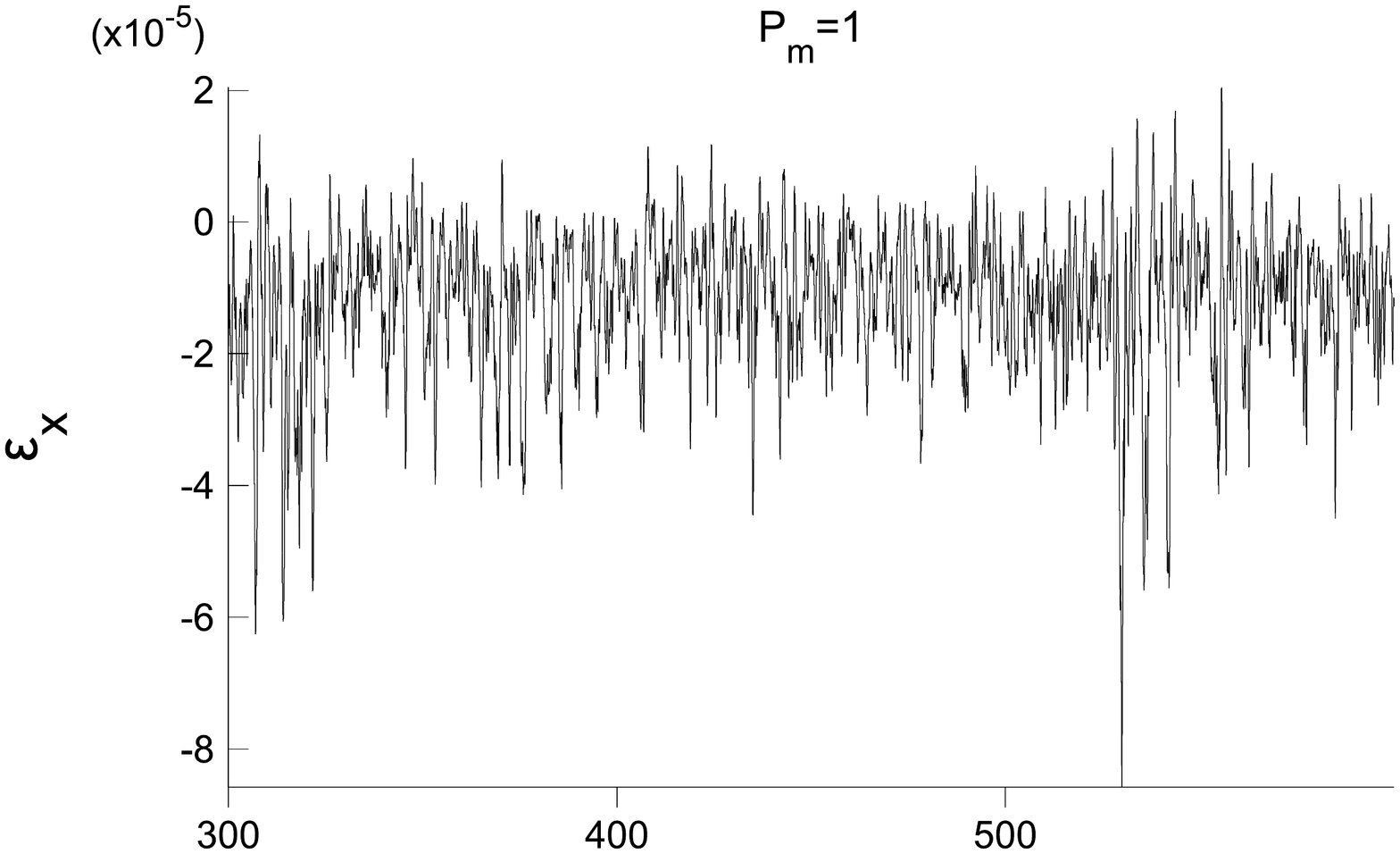} 
\par\end{centering}

\begin{centering}
\includegraphics[width=.6\linewidth,angle=-90]{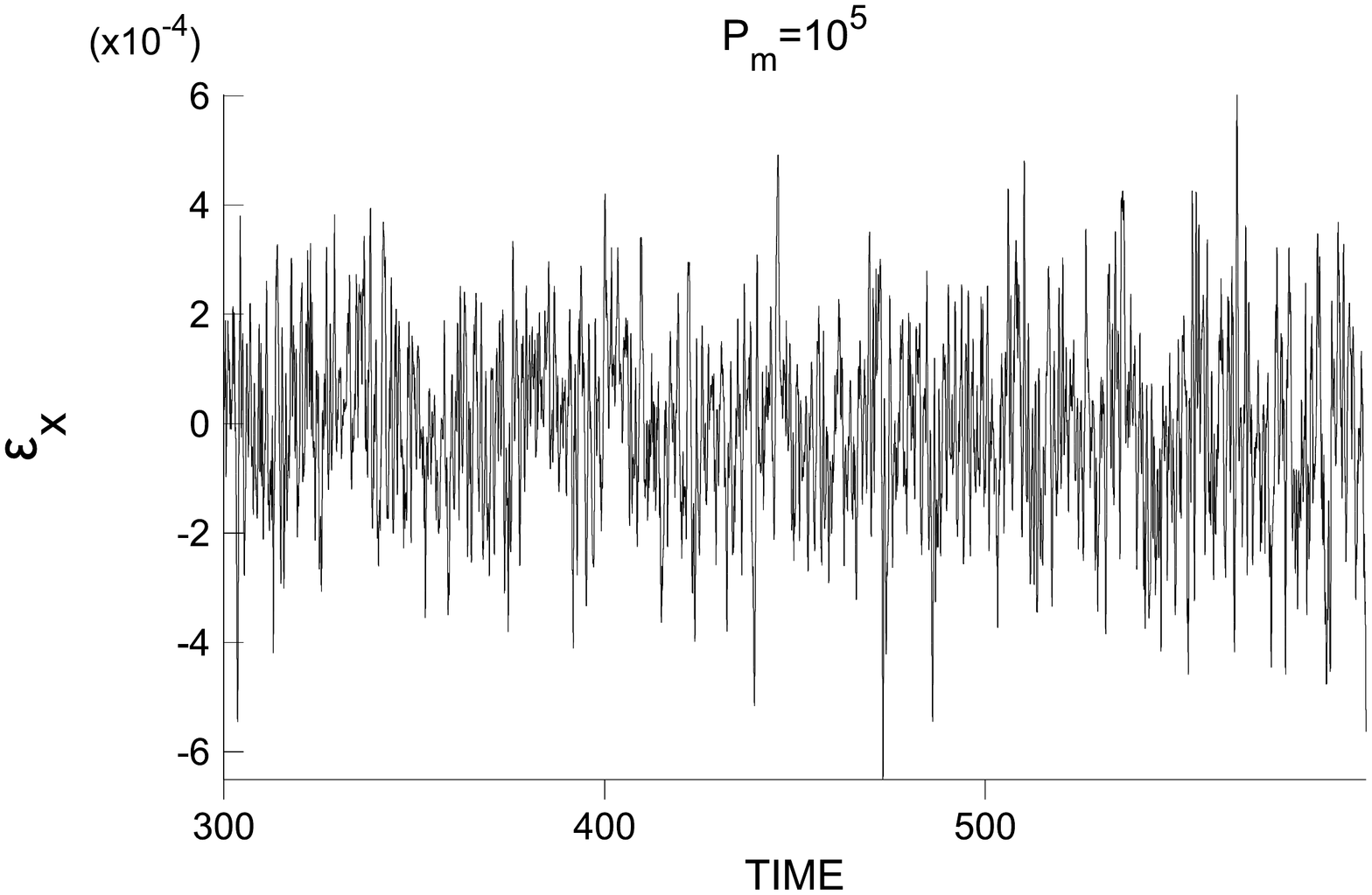} 
\par\end{centering}

\caption{\label{f1}Electromotive force in $x$ direction for the low(top)
and high(bottom) $P_{m}$, $\overline{B}/\nu=0.1$. }

\end{figure}

The shell-model approach has been widely used in turbulence modelling \citep{GledDolOb:81,Bohr98}.
A combination of the mean-field dynamo with a shellmodel was explored
in \cite{SokFr03}. There a dynamical system based on the shell-model
was invoked to describe the dynamics of the small-scale fluctuating
velocity and magnetic fields. Here, we utilize a similar idea but
with a different purpose. Consider a velocity field with the Fourier
representation \[
\mathbf{u}\left(\mathbf{x}\right)=\sum_{n=0}^{N}\left(\mathbf{\hat{u}}^{\left(n\right)}\mathbf{e}^{\imath\left(\mathbf{k}^{(n)}\cdot\mathbf{x}\right)}+\widetilde{\widehat{\mathbf{u}}}^{\left(n\right)}\mathbf{e}^{-\imath\left(\mathbf{k}^{(n)}\cdot\mathbf{x}\right)}\right).\]
 Let $N=4$ and the wave-vectors form a tetrahedron as shown in Figure
\ref{fig1}. Without loss of generality the wavevectors may be taken
to have unit modulus. We suppose that the fluctuating magnetic field
has the same representation, and that the nonlinear coupling terms
only project onto this same set of vectors. It may be shown that the
resulting closed nonlinear system obeys all the usual conservation
laws in the absence of diffusion, and so seems a useful test bed for
examining the accuracy of the various approximations. Projecting equations
(\ref{eq:1a},\ref{eq:1b}) onto the given Fourier components we get
equations for the modes: \begin{eqnarray}
\partial_{t}\widehat{\mathbf{b}}^{\left(l\right)} & = & -\eta\widehat{\mathbf{b}}^{\left(l\right)}+\imath\left(\mathbf{\overline{B}}\cdot\mathbf{k}^{\left(1\right)}\right)\mathbf{\hat{u}}^{\left(1\right)}+\mathbf{\mathcal{M}}^{\left(l\right)}-\mathbf{\overline{\mathcal{M}}}^{\left(l\right)},\label{eq:1}\\
\partial_{t}\widehat{\mathbf{u}}^{\left(l\right)} & = & -\nu\widehat{\mathbf{u}}^{\left(l\right)}+\imath\left(\mathbf{\overline{B}}\cdot\mathbf{k}^{\left(1\right)}\right)\mathbf{\hat{b}}^{\left(1\right)}\label{eq:2}\\
 & + & \pi^{\left(l\right)}\circ\left(\mathbf{\mathcal{N}}^{\left(l\right)}-\overline{\mathbf{\mathcal{N}}}^{\left(l\right)}\right)+\pi^{\left(l\right)}\circ\mathbf{f}^{\left(l\right)}\nonumber \end{eqnarray}
 where the superscript $^{(l)}$ means the number of the mode and
$\pi_{ij}^{(l)}=\delta_{ij}-k_{i}^{(l)}k_{j}^{(l)}$. The nonlinear
contributions are given in terms of the tensors $\mathbf{\mathcal{M}}^{\left(l\right)}$
and $\pi^{\left(l\right)}\circ\mathbf{\mathcal{N}}^{\left(l\right)}$
which are shown in Appendix A. We suppose for simplicity that $\nabla\cdot\mathbf{b}=\nabla\cdot\mathbf{u}=0$
so that each modal equation has all its terms perpendicular to ${\mathbf{k}}^{(l)}$.
Equations (\ref{eq:1},\ref{eq:2}) will be solved numerically.
The FOSA solutions correspond to the case where all nonlinear contributions
in (\ref{eq:1},\ref{eq:2}) are neglected.

To formulate the variant of the $\tau$-approximation which is relevant
for the given model we need equations for the second-order products
of the fluctuating fields averaged over the ensemble of fluctuations.
Starting from (\ref{eq:1},\ref{eq:2}) we get: 
{\small
\begin{eqnarray}
\partial_{t}\left(\overline{\widehat{b_{i}}^{\left(l\right)}\widetilde{\hat{b}}_{j}^{\left(l\right)}}\right) & = & -2P_{m}^{-1}\overline{\widehat{b_{i}}^{\left(l\right)}\widetilde{\hat{b}}_{j}^{\left(l\right)}}+\nu^{-1}\left(\overline{\widetilde{\mathcal{M}}_{j}^{\left(l\right)}\hat{b}_{i}^{\left(l\right)}}+\overline{\mathcal{M}_{i}^{\left(l\right)}\widetilde{\hat{b}}_{j}^{\left(l\right)}}\right)\nonumber\\
 & + & \nu^{-1}\left(\mathbf{\overline{B}}\cdot\mathbf{k}^{\left(1\right)}\right)\left(\overline{\widehat{u_{i}}^{\left(l\right)}\widetilde{\hat{b}}_{j}^{\left(l\right)}}-\overline{\widehat{b_{i}}^{\left(l\right)}\widetilde{\hat{u}}_{j}^{\left(l\right)}}\right),\label{eq:bt} \\
\partial_{t}\left(\overline{\widehat{u_{i}}^{\left(l\right)}\widetilde{\hat{u}}_{j}^{\left(l\right)}}\right) & = & -2\overline{\widehat{u_{i}}^{\left(l\right)}\widetilde{\hat{u}}_{j}^{\left(l\right)}}\nonumber\\
&-&\nu^{-1}\left(\mathbf{\overline{B}}\cdot\mathbf{k}^{\left(1\right)}\right)\left(\overline{\widehat{u_{i}}^{\left(l\right)}\widetilde{\hat{b}}_{j}^{\left(l\right)}}-\overline{\widehat{b_{i}}^{\left(l\right)}\widetilde{\hat{u}}_{j}^{\left(l\right)}}\right)\nonumber\\
 &+& \nu^{-1}\left(\overline{\widetilde{\mathcal{N}}_{j}^{\left(l\right)}\hat{u}_{i}^{\left(l\right)}}+\overline{\mathcal{N}_{i}^{\left(l\right)}\widetilde{\hat{u}}_{j}^{\left(l\right)}}\right.\nonumber\\
 &\phantom{+}&~~~~~+~\left.\overline{\widetilde{f^{(s)}}_{j}^{\left(l\right)}\hat{u}_{i}^{\left(l\right)}}+\overline{f_{i}^{(s)\left(l\right)}\widetilde{\hat{u}}_{j}^{\left(l\right)}}\right),\label{eq:ut} \end{eqnarray}
 \begin{eqnarray}
\partial_{t}\left(\overline{\widehat{u_{i}}^{\left(l\right)}\widetilde{\hat{b}}_{j}^{\left(l\right)}}\right)\!\! & = & \!\!-\left(1+P_{m}^{-1}\right)\overline{\widehat{u_{i}}^{\left(l\right)}\widetilde{\hat{b}}_{j}^{\left(l\right)}} + \nu^{-1}\overline{f_{i}^{(s)\left(l\right)}\widetilde{\hat{b}}_{j}^{\left(l\right)}}\nonumber\\
&+&\nu^{-1}\left(\mathbf{\overline{B}}\cdot\mathbf{k}^{\left(1\right)}\right)\left(\overline{\widehat{b_{i}}^{\left(l\right)}\widetilde{\hat{b}}_{j}^{\left(l\right)}}\!\!-\overline{\widehat{u_{i}}^{\left(l\right)}\widetilde{\hat{u}}_{j}^{\left(l\right)}}\right)\nonumber\\
 &+&\nu^{-1}\left(\overline{\tilde{M}_{j}^{\left(l\right)}\hat{u}_{i}^{\left(l\right)}}+\overline{\mathcal{N}_{i}^{\left(l\right)}\widetilde{\hat{b}}_{j}^{\left(l\right)}}\right),\label{eq:etau}
\end{eqnarray}}
where the tilde above physical quantities means the complex conjugate
and averaging over the ensemble of fluctuations is denoted by an
overbar. In the $\tau$-approximation (see, e.g. \cite{bra-sub:04,kle-rog:04a}
) we replace the third order contributions in (\ref{eq:bt},\ref{eq:ut},\ref{eq:etau})
by the corresponding relaxation terms of the second-order contributions.
For example, in (\ref{eq:etau}) we set
\begin{equation}
\nu^{-1}\left(\overline{\tilde{M}_{j}^{\left(l\right)}\hat{u}_{i}^{\left(l\right)}}+\overline{\mathcal{N}_{i}^{\left(l\right)}\widetilde{\hat{b}}_{j}^{\left(l\right)}}\right)=-\tau^{-1}\overline{\widehat{u_{i}}^{\left(l\right)}\widetilde{\hat{b}}_{j}^{\left(l\right)}},\label{eq:tauap}\end{equation}
 where $\tau$ denotes the typical relaxation time of the fluctuating
terms. In this formulation $\tau$ is an external parameter of this approximation.
We do not need to solve equations (\ref{eq:bt},\ref{eq:ut},\ref{eq:etau}).
Instead we will use the left part of (\ref{eq:tauap}) to find the
mean electromotive force obtained with the $\tau$-approximation.

\begin{figure}
\begin{centering}
\includegraphics[height=.6\linewidth,angle=-90]{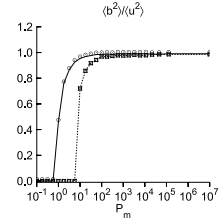} 
\par\end{centering}

\caption{\label{f2}Relation between the energy of the small-scale velocity
and magnetic fields. Squares are for Case 1 and circles are for Case 2.}

\end{figure}

\begin{figure}
\begin{centering}
\includegraphics[height=.5\linewidth,angle=-90]{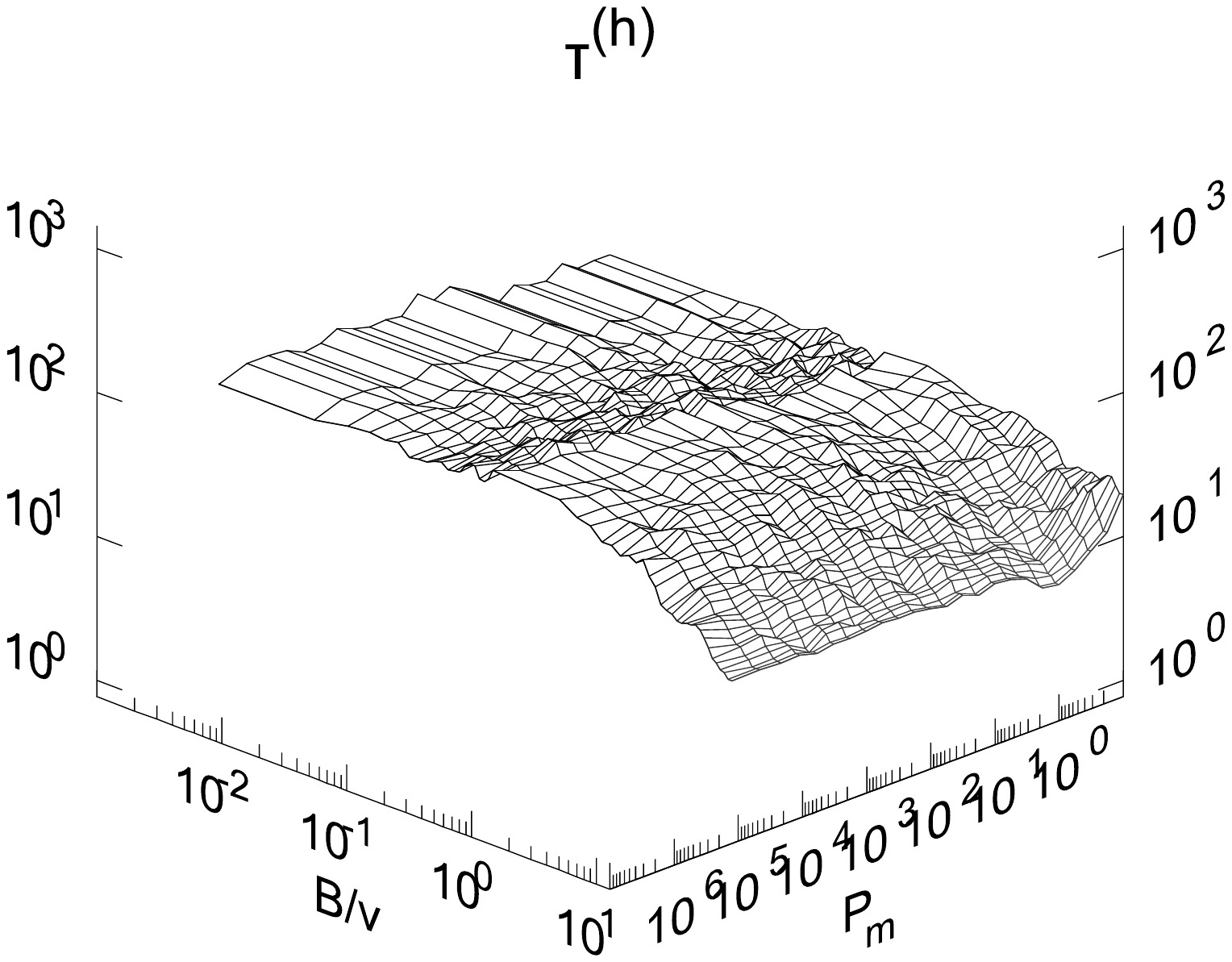}\includegraphics[height=.5\linewidth,angle=-90]{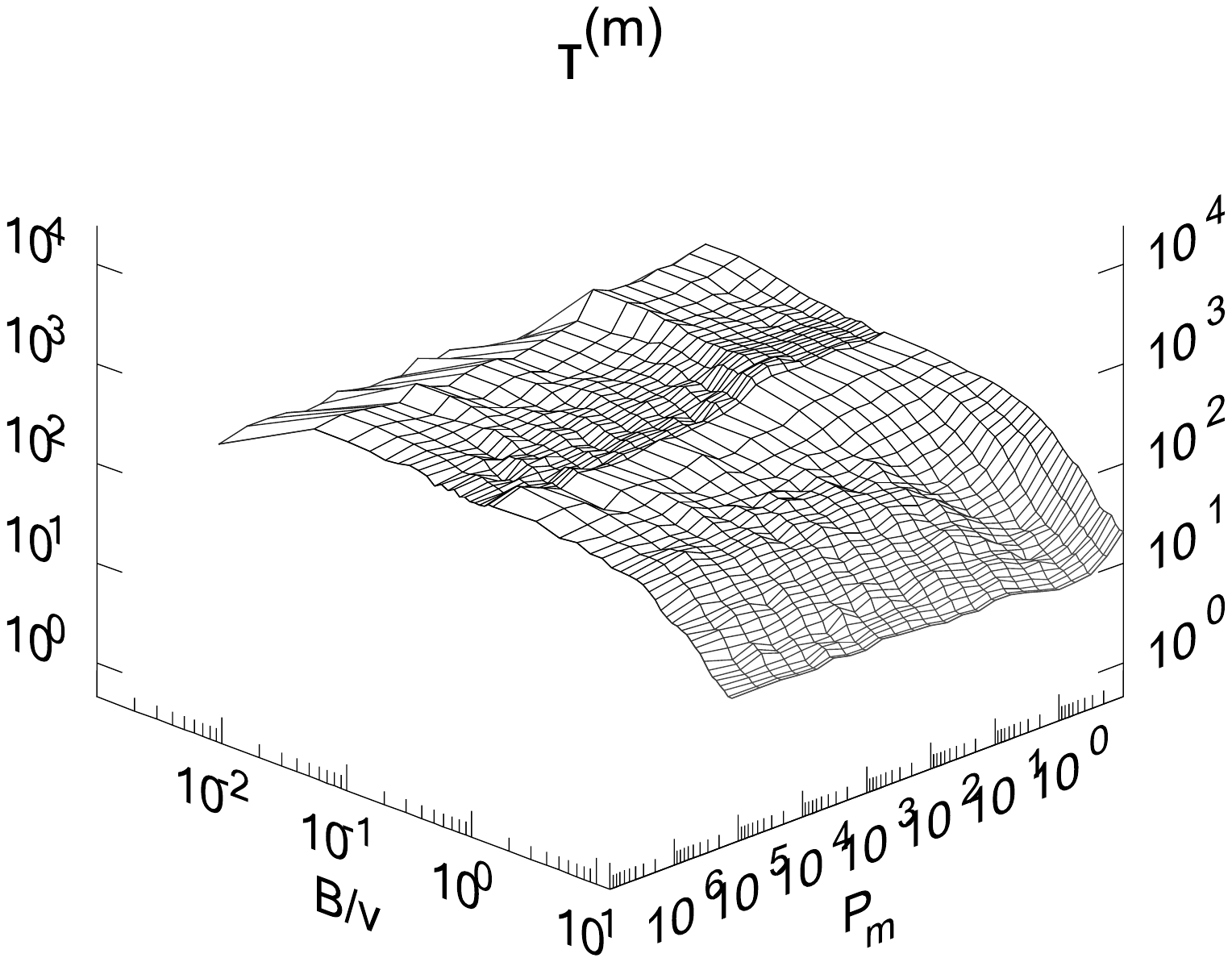} 
\par\end{centering}

\begin{centering}
\includegraphics[height=.5\linewidth,angle=-90]{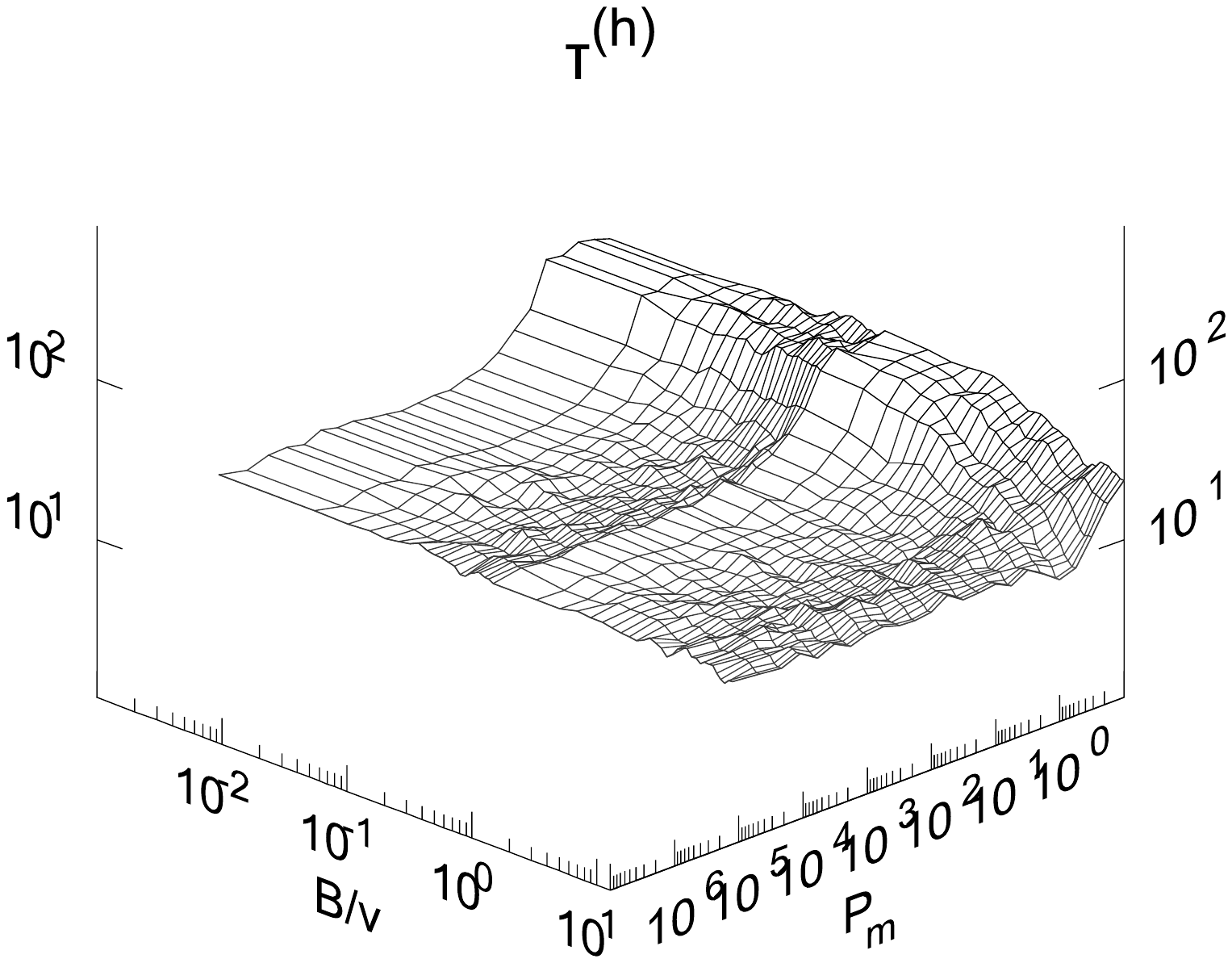}\includegraphics[height=.5\linewidth,angle=-90]{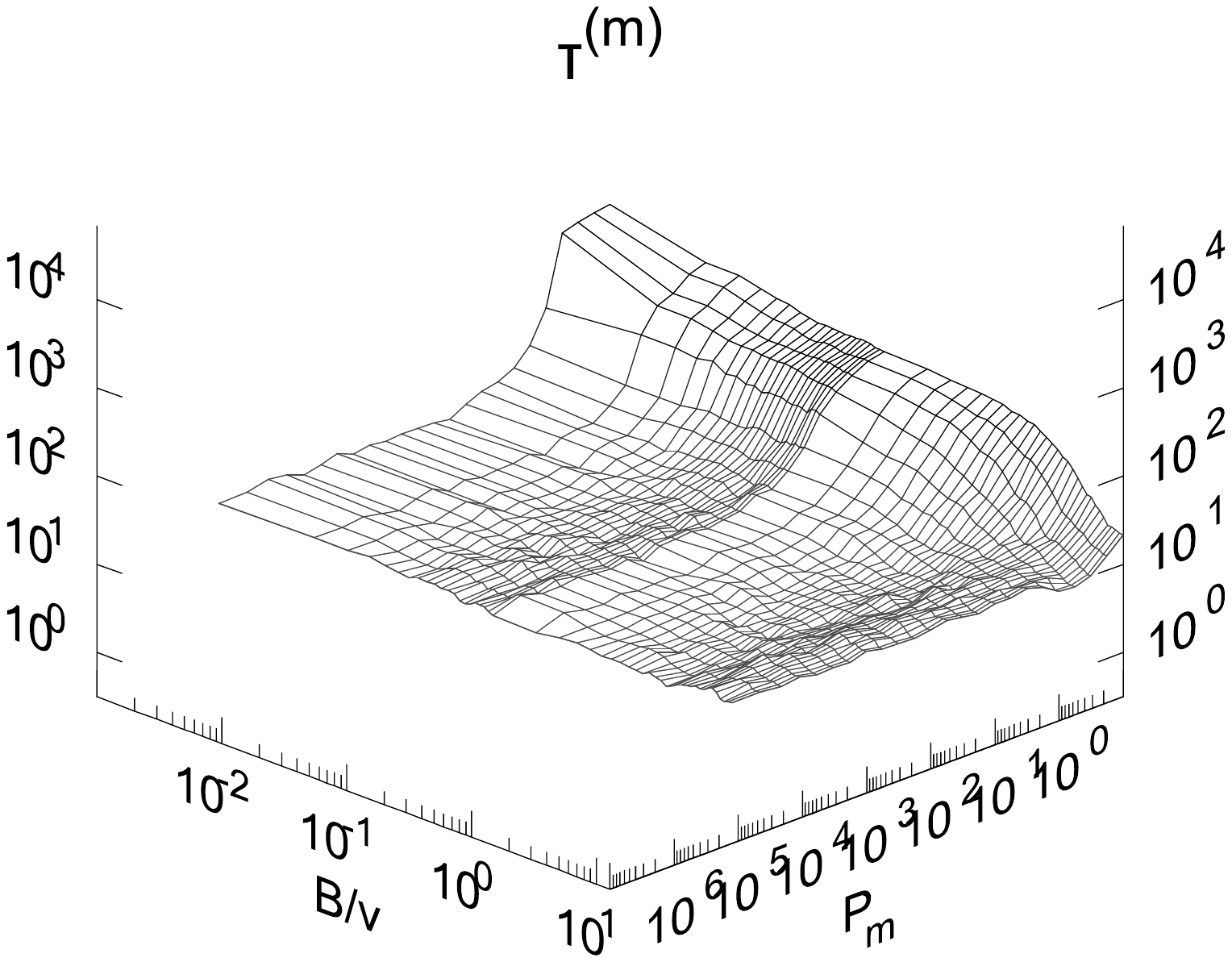} 
\par\end{centering}

\caption{\label{fig:taus}Correlation time. Left - hydrodynamic, right - magnetic.
Upper panel: Case 1; lower panel: Case 2 }

\end{figure}

The model (\ref{eq:1},\ref{eq:2}) is clearly a good one when the
diffusivities are large, but will not give any better results than
the other truncations when the diffusivities are small. Nonetheless it
does provide a useful simplification in mid ranges and permits the
testing of the various approximations. Plainly a major simplification
is that the fields are monochromatic. This could and should be remedied
by increasing the number of shells, but this has not yet been attempted.

\section*{The model design}

Equations (\ref{eq:1}, \ref{eq:2}) were solved numerically using
a second order time integration scheme. Time is measured by the typical
diffusion time, $t\rightarrow t/\nu$. The random force is normalized
with $\nu$ as well, $\mathbf{f}\rightarrow\mathbf{f}\nu$.

The evolution of the small-scale velocity and magnetic fields depends
on the typical correlation time of the random force. The time
step is $0.003$ (in dimensionless units, time is rescaled according to
$t\rightarrow\nu t$).  In what follows we consider two different
cases. Case 1 is that of zero correlation time: the force is updated
at each timestep. In Case 2, which has finite correlation time, the
force was updated each 50-th time step. 

The effective Reynolds number is given by $Re=u_{c}\ell_{c}/\nu$.
In computations presented below we use $\nu=0.05$ and $\nu=0.01$.
We define the random driving force by writing $\mathbf{f}^{l}=\mathbf{w}^{(l)}+i\mathbf{k}^{(l)}\times\mathbf{w}^{(l)}$,
and similar for initial velocity and magnetic fields. For each $l$
$\mathbf{w}^{(l)}$ is a random vector whose components vary between
$\pm0.5$. The term $i\mathbf{k}^{(l)}\times\mathbf{w}^{(l)}$ is
introduced to force positive helicity in the system. The initial velocity
field is given helicity of the same sign. The electromotive force
associated with the $\left(l\right)-$mode reads \begin{eqnarray}
\mathcal{E}_{i}^{(l)} & = & \varepsilon_{ijq}\hat{u}_{j}^{(l)}\widetilde{\hat{b}}_{q}^{(l)}~+c.c.~=\varepsilon_{ijq}\chi_{jq}^{(l)}\label{eq:emff}\end{eqnarray}
 where tilde means the complex conjugate. Suppose the mean magnetic
field has fixed direction, $\overline{\mathbf{B}}=\mathbf{e}_{x}B_{x}$.
The important component of the mean electromotive force is $\mathcal{E}_{x}$,
and so we define the $\alpha$ effect via $\alpha={\overline{\mathcal{E}}}_{x}/B_{x}$.
The mean electromotive force ${\overline{\mathbf{\mathcal{E}}}}$
is found by summation over all modes and in averaging over the long-time
interval equal to about 3000 diffusion times of the system (here $M$
is the total number of time-steps) \begin{equation}
\overline{\mathcal{E}}_{i}=\varepsilon_{ijq}\frac{1}{M}\sum_{m=0}^{m=M}\sum^{(l)}\chi_{jq}^{(l)}\end{equation}
 Typical realizations of $\mathcal{E}_{x}=\varepsilon_{ijq}\sum^{(l)}\chi_{jq}^{n}$
for case $P_{m}=\infty$ and $P_{m}=1$ are shown on the figure \ref{f1}.
The averaging was done over 16 such realizations. For the purpose
of comparison we also solve equations (\ref{eq:1}, \ref{eq:2}) using
the first order smoothing approximation (FOSA), in which the tensors
$\mathcal{M},\mathcal{N}$ are set to zero. To test the $\tau$ approximation
we evaluate the third order moments (see explanations above): \begin{equation}
\chi_{ij}^{(l),\ \tau}=\nu^{-1}\left(\overline{\tilde{M}_{j}^{\left(l\right)}\hat{u}_{i}^{\left(l\right)}}+\overline{\mathcal{N}_{i}^{\left(l\right)}\widetilde{\hat{b}}_{j}^{\left(l\right)}}\right)\label{eq:emftau}\end{equation}

\begin{figure}
\begin{centering}
\includegraphics[height=.5\linewidth,angle=-90]{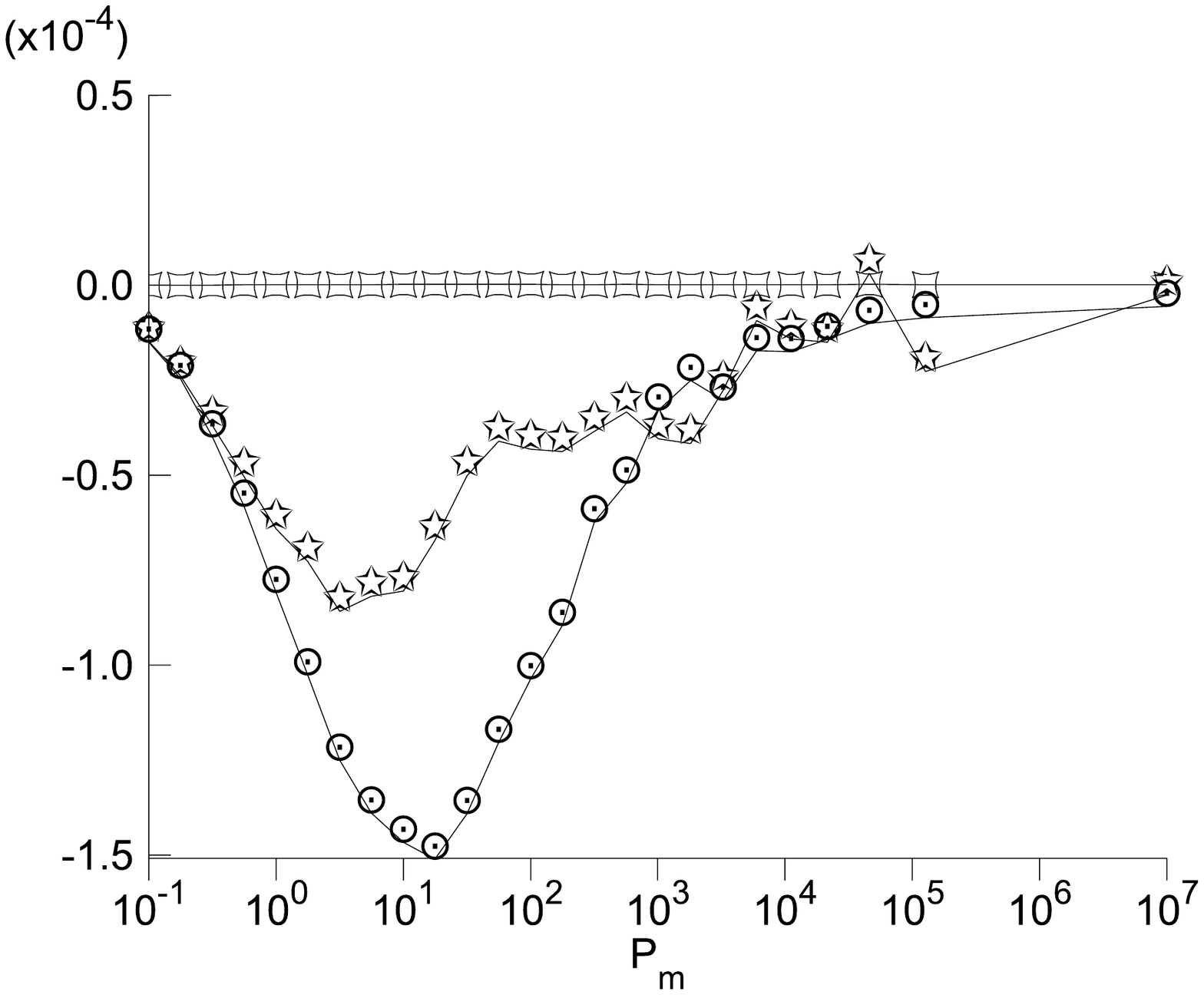}\includegraphics[height=.5\linewidth,angle=-90]{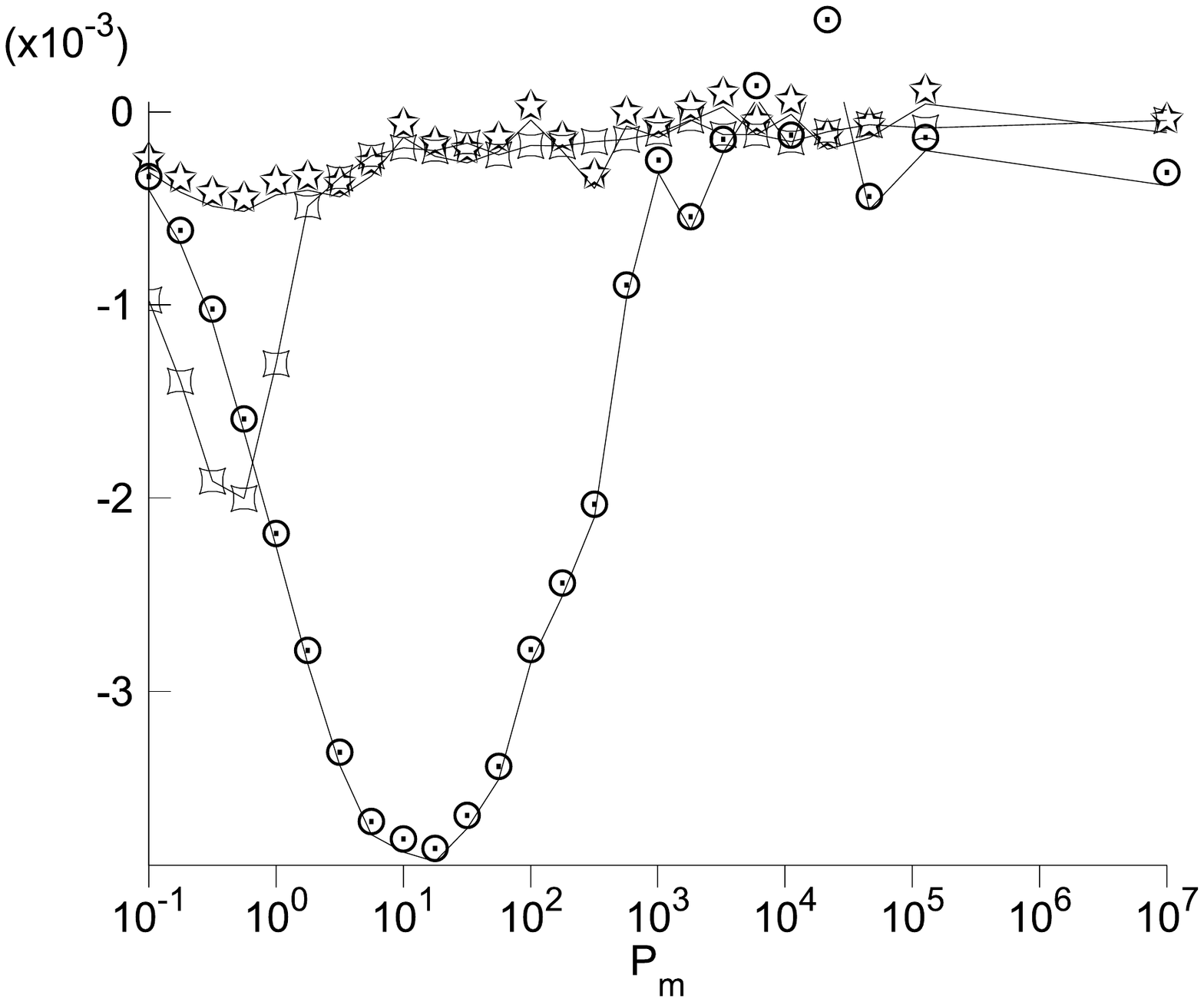} 
\par\end{centering}

\caption{\label{fig:EPm} $\mathcal{E}$ vs $P_{m}$, $B/\nu=0.1$. Left: Case
1. The FOSA solution is shown by circles, the exact solution by stars
and squares show the $\tau$ approximation. The right hand graph shows the same
data for Case 2.}

\end{figure}

\begin{figure}
\begin{centering}
\includegraphics[width=.5\linewidth,angle=-90]{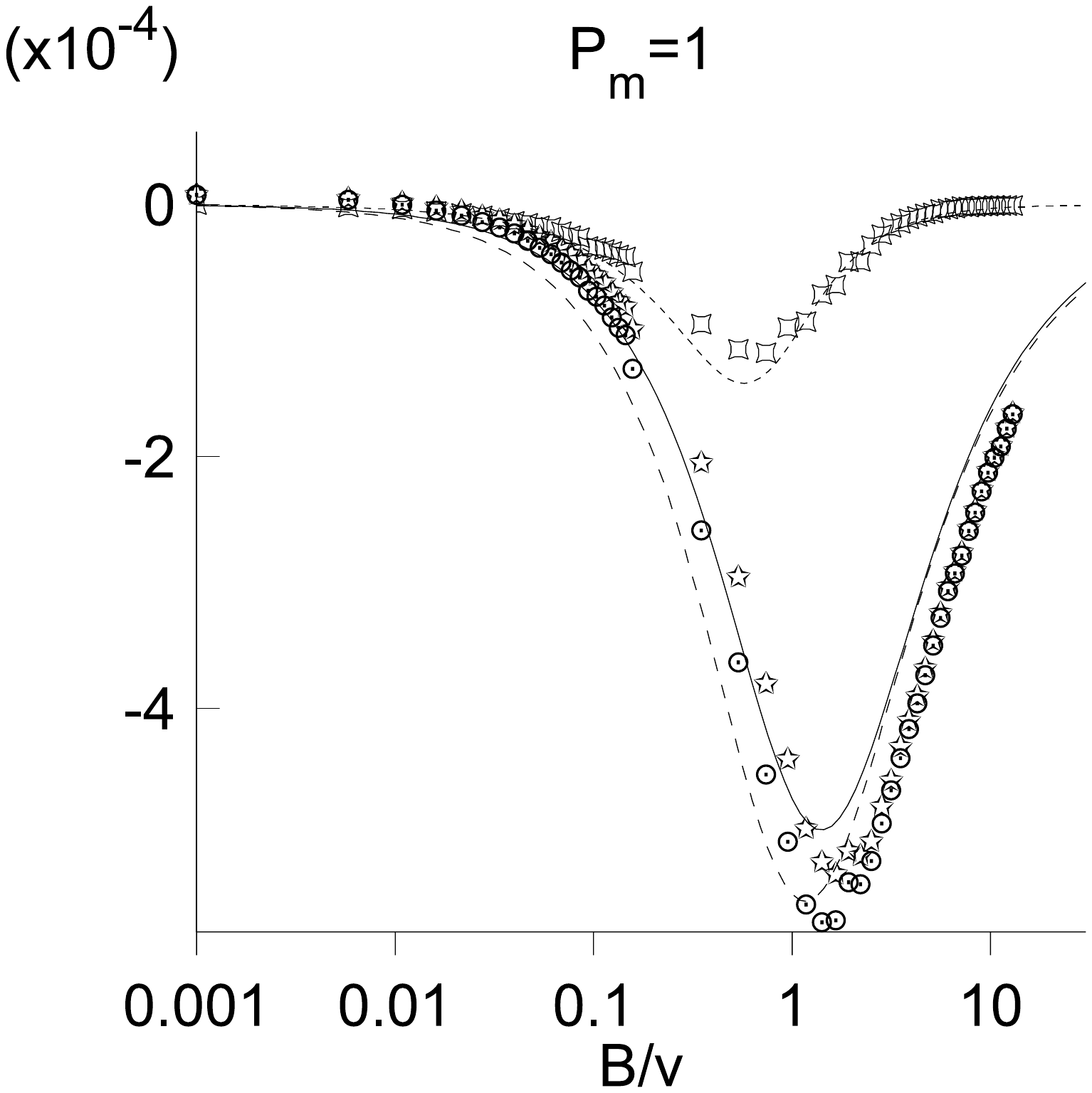}\includegraphics[width=.5\linewidth,angle=-90]{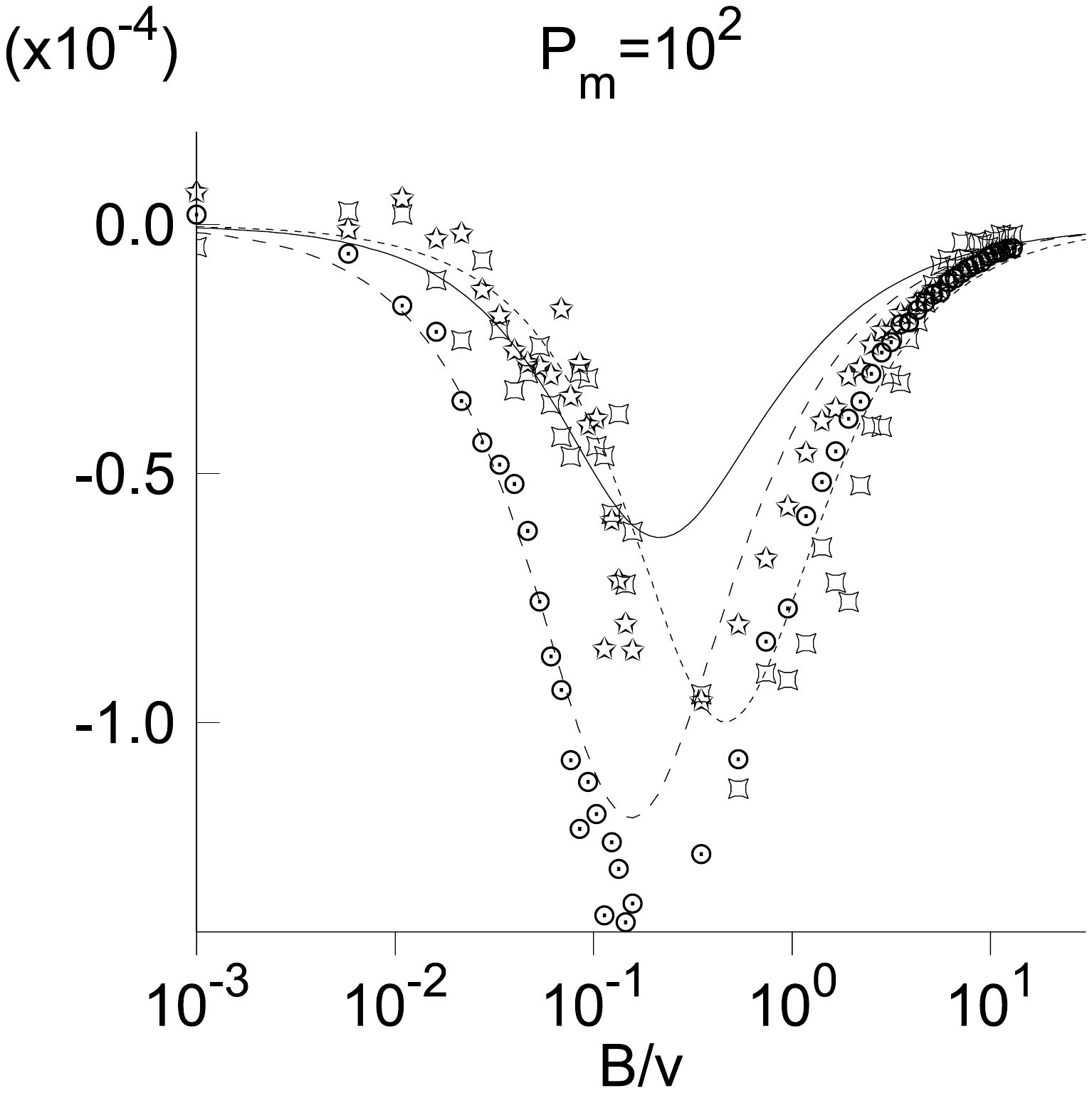} 
\par\end{centering}

\vspace{0.25cm}

\begin{centering}
\includegraphics[width=.5\linewidth,angle=-90]{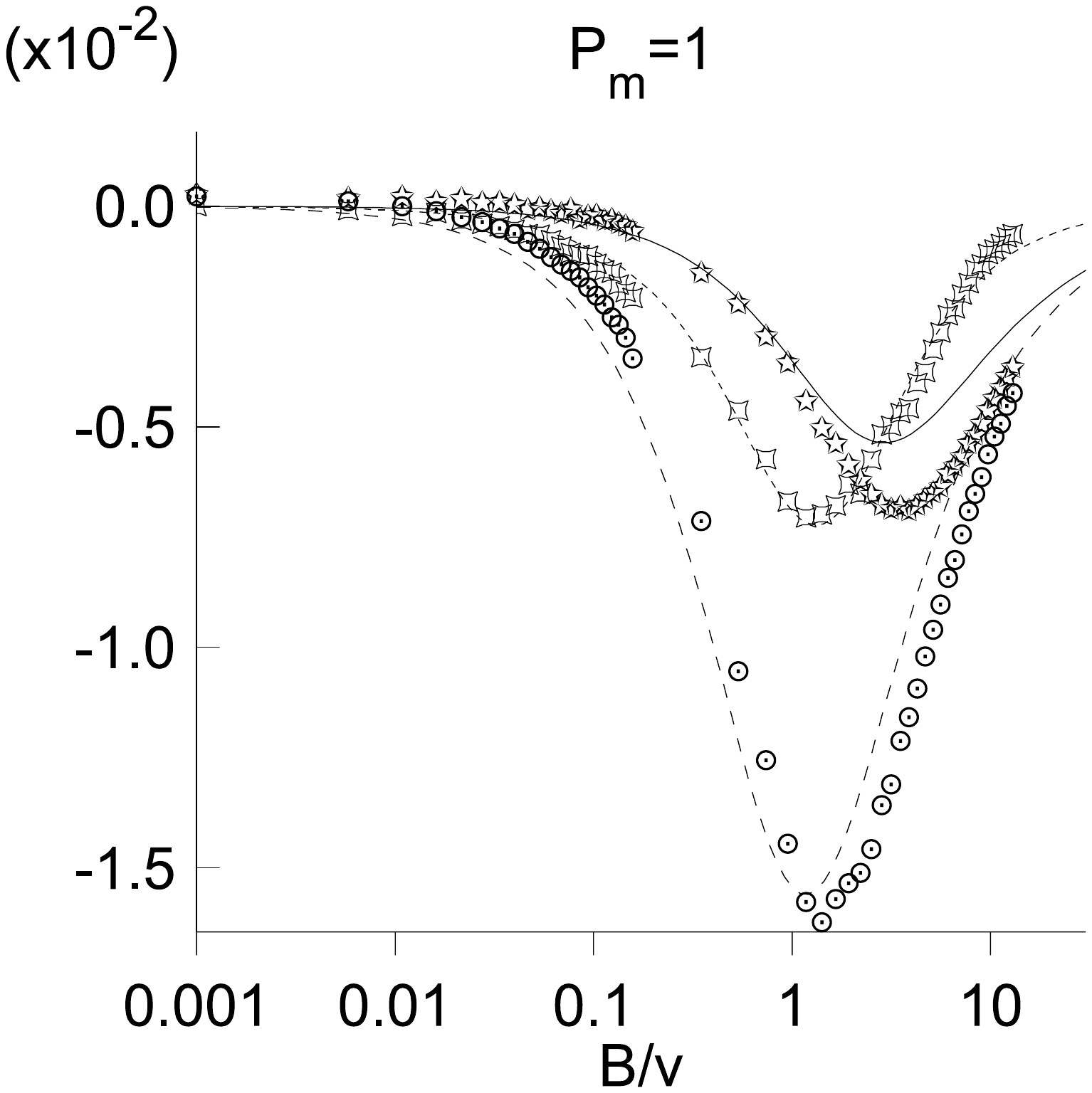}\includegraphics[width=.5\linewidth,angle=-90]{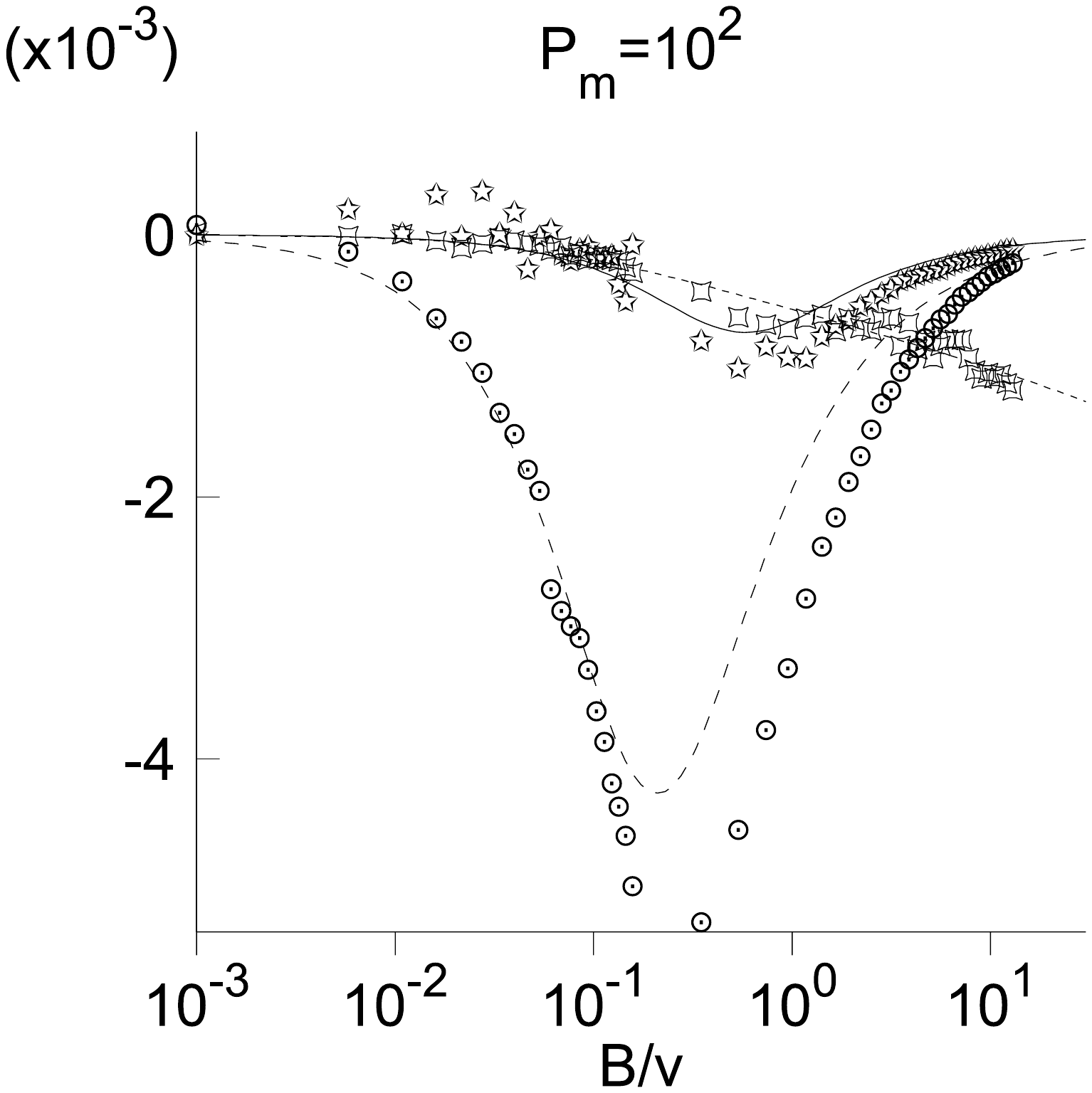} 
\par\end{centering}

\caption{\label{fig:fits}$\mathcal{E}_{x}$ vs $B/\nu$. The top row shows
comparisons between the full solution of the model and the approximations
for Case 1. At left - the case $P_{m}=1$ at right - $P_{m}=10^{3}$.
The FOSA solution is shown by circles, the full solution by stars,
and squares show the $\tau$ approximation. The bottom shows the same
data for Case 2. }

\end{figure}

First we give a detailed description of results for the case $\nu=0.05$.
The results depend very much on whether there is a small-scale dynamo
- that is whether a small scale field can exist in the absence of
the imposed large scale field. The value of $P_{m}$ affects both
the threshold and intensity of the small-scale dynamo. The relation
between $P_{m}$ and the amplitude of the small-scale magnetic field
fluctuations for $\overline{B}=0$ is shown in Figure \ref{f2}. For
Case 1 there is a small-scale dynamo if $P_{m}\ge10$.  Also, as may be verified directly from the equations, the amplitude
of the mean electro-motive force tends to zero if $P_{m}$ approaches
 infinity. This is illustrated in Figure \ref{fig:EPm} for the
case $B/\nu=1.$ The typical Reynolds number is $Re\approx2.2$ for
 Case 1 and $Re\approx4.8$ for Case 2.

For Case 2 the threshold is about, $P_{m}\approx1$. We can 
see that for large $P_{m}$ there is approximate equipartition between the energies
of the fluctuating velocity and magnetic field. 

As well as examining
the accuracy of FOSA, we will explore the usfulness of the $\tau$-approximation.
The approximation relies on knowledge of the typical relaxation times
$\tau^{(m)},\tau^{(h)}$ of magnetic and hydrodynamic fluctuations.
These quantities were found from auto-correlation functions; for $\tau^{(h)}$
we have \begin{eqnarray*}
\mathcal{I}^{\left(h\right)}\left(T\right) & = & \frac{\overline{\int u_{i}^{\left(l\right)}\left(t\right)u_{i}^{\left(l\right)}\left(t+T\right)dt}}{\overline{\int u_{i}^{\left(l\right)}\left(t\right)u_{i}^{\left(l\right)}\left(t\right)dt}},\\
\tau^{\left(h\right)} & =T & \left(\frac{\mathcal{I}^{\left(h\right)}\left(0\right)}{\mathcal{I}^{\left(h\right)}\left(T\right)}=e^{-1}\right),\end{eqnarray*}
 and similarly for $\tau^{\left(m\right)}$. Both $\tau^{\left(h\right)}$
and $\tau^{\left(m\right)}$ are functions of $\overline{B}$ and
$P_{m}$, as shown in Figure \ref{fig:taus}. As we can see, $\tau^{\left(h\right)}$
depends strongly on $\overline{B}$. Its variation with $P_{m}$ depends
on the existence of the small-scale dynamo. On the other hand, $\tau^{\left(m\right)}$
does not show considerable variation either with $\overline{B}$ or
$P_{m}$. To estimate the results of $\tau$ approximation we take
$\overline{\tau}=\left(\tau^{\left(h\right)}+\tau^{\left(m\right)}\right)/2$.

The dependence of the calculated mean electromotive force on $P_{m}$
for the fixed strength of $\overline{B}$ is shown in Figure\ref{fig:EPm}.
The maxima are at values of $P_{m}$ that are close to the thresholds for the small-scale
dynamo. For high $P_{m}$,  $\mathcal{E}$  fluctuates strongly
about zero. The dependence of the magnitude of the mean electromotive force on  $P_{m}$
is not easily determined for small values of  $\overline{B}$ because of strong fluctuations.

\begin{figure}
\begin{centering}
\includegraphics[width=.32\linewidth,angle=-90]{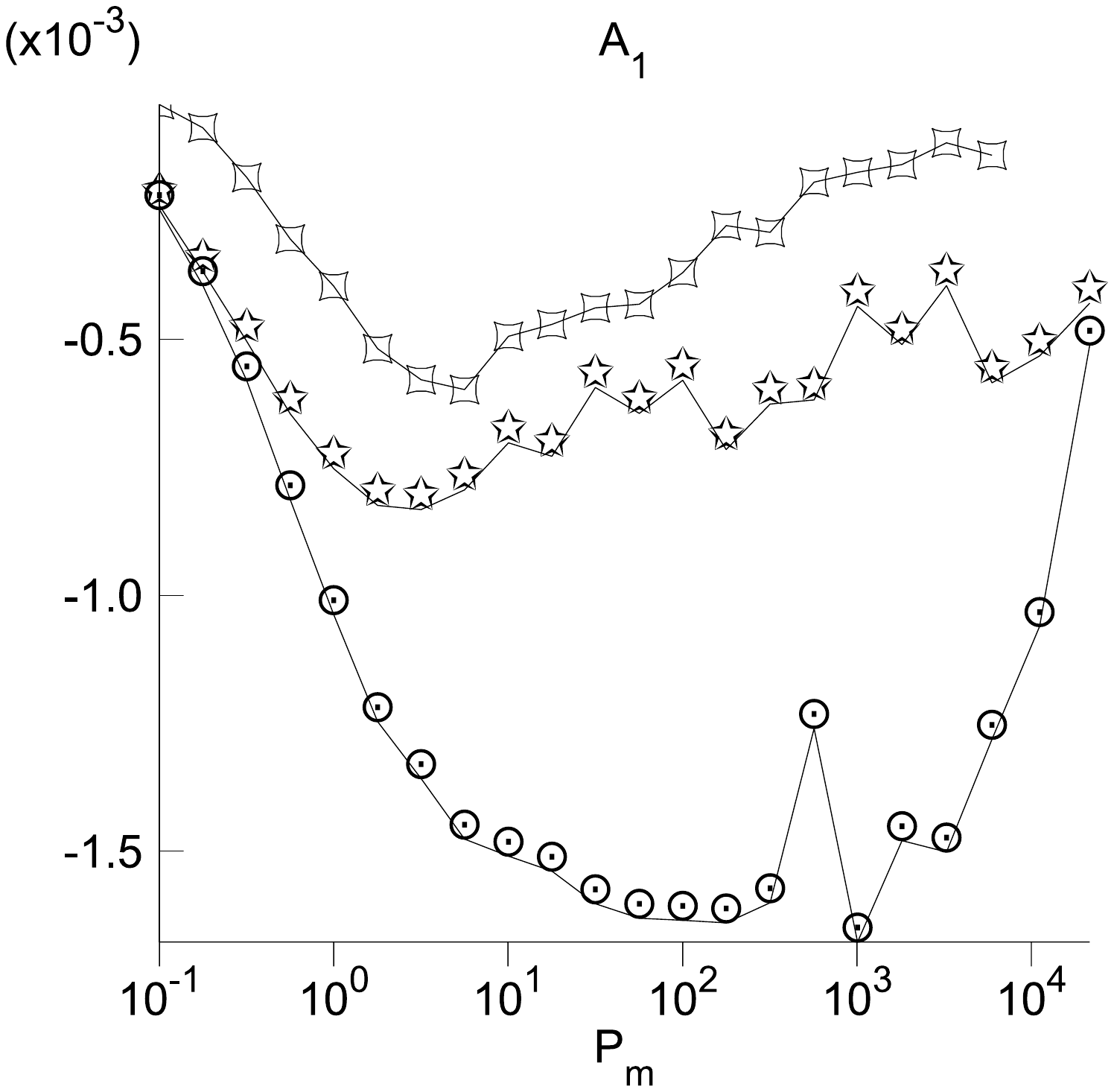}\ \ \includegraphics[width=.32\linewidth,angle=-90]{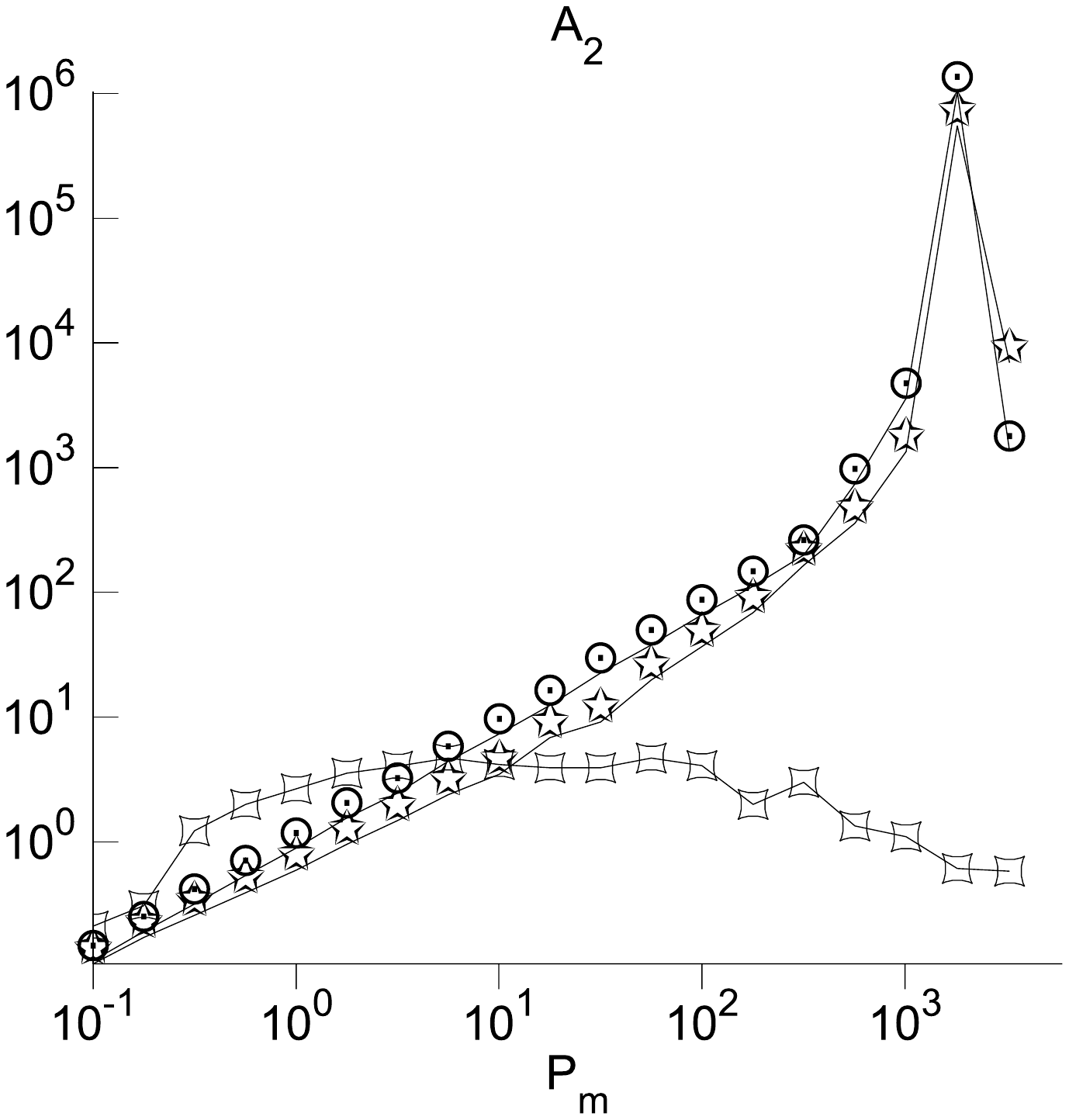}\ \ \includegraphics[width=.32\linewidth,angle=-90]{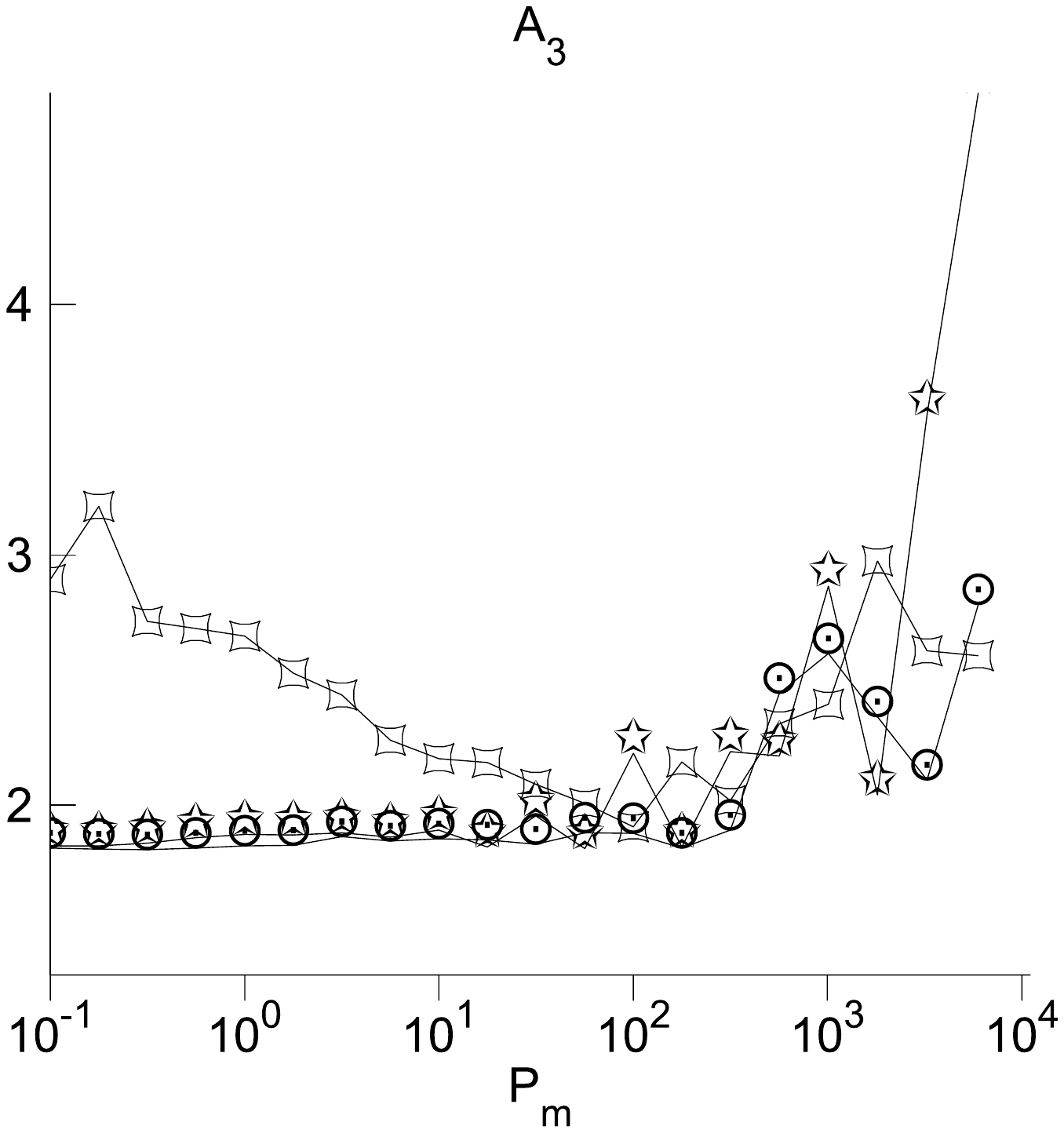} 
\par\end{centering}

\begin{centering}
\includegraphics[width=.32\linewidth,angle=-90]{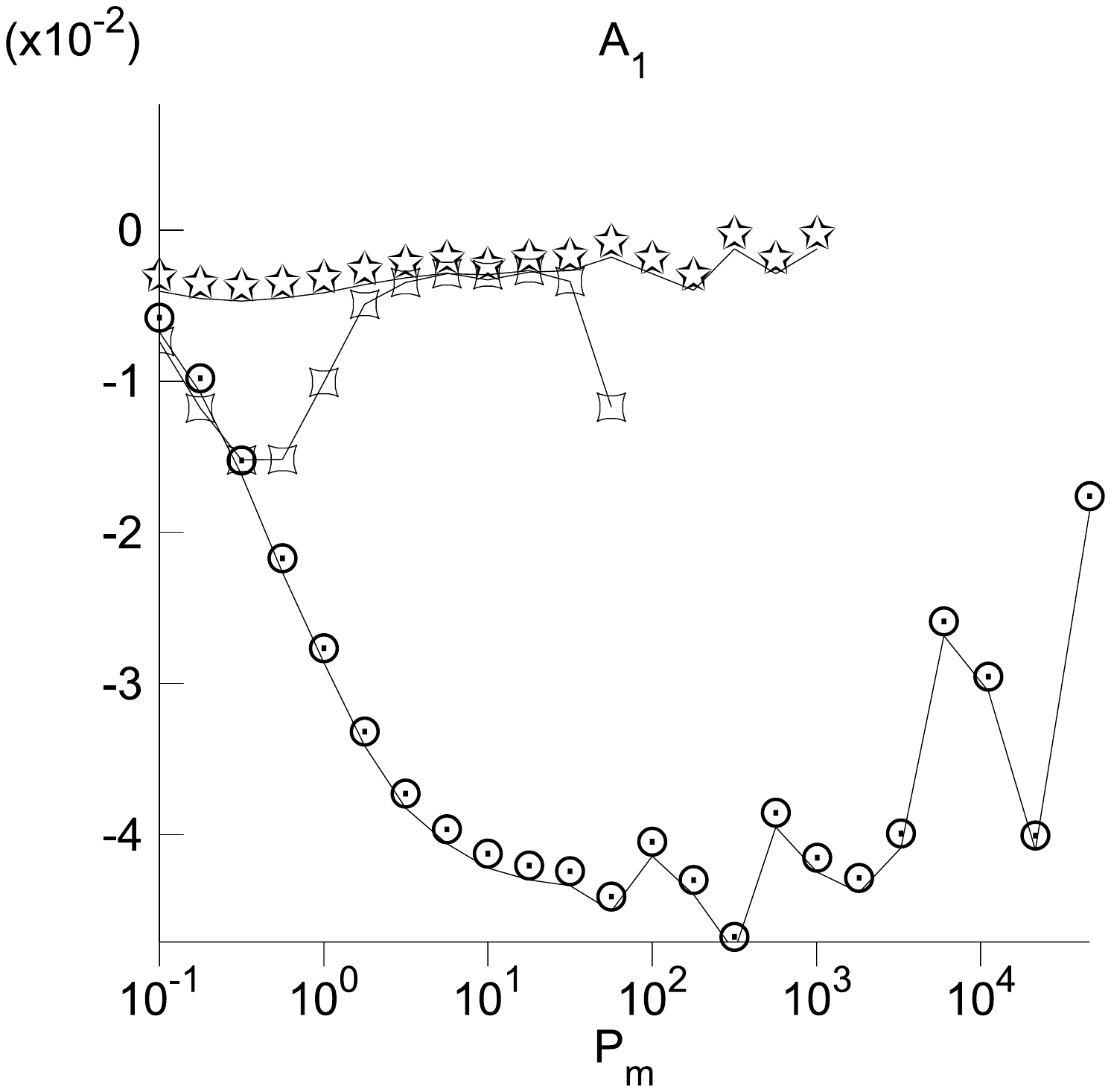}\ \ \includegraphics[width=.32\linewidth,angle=-90]{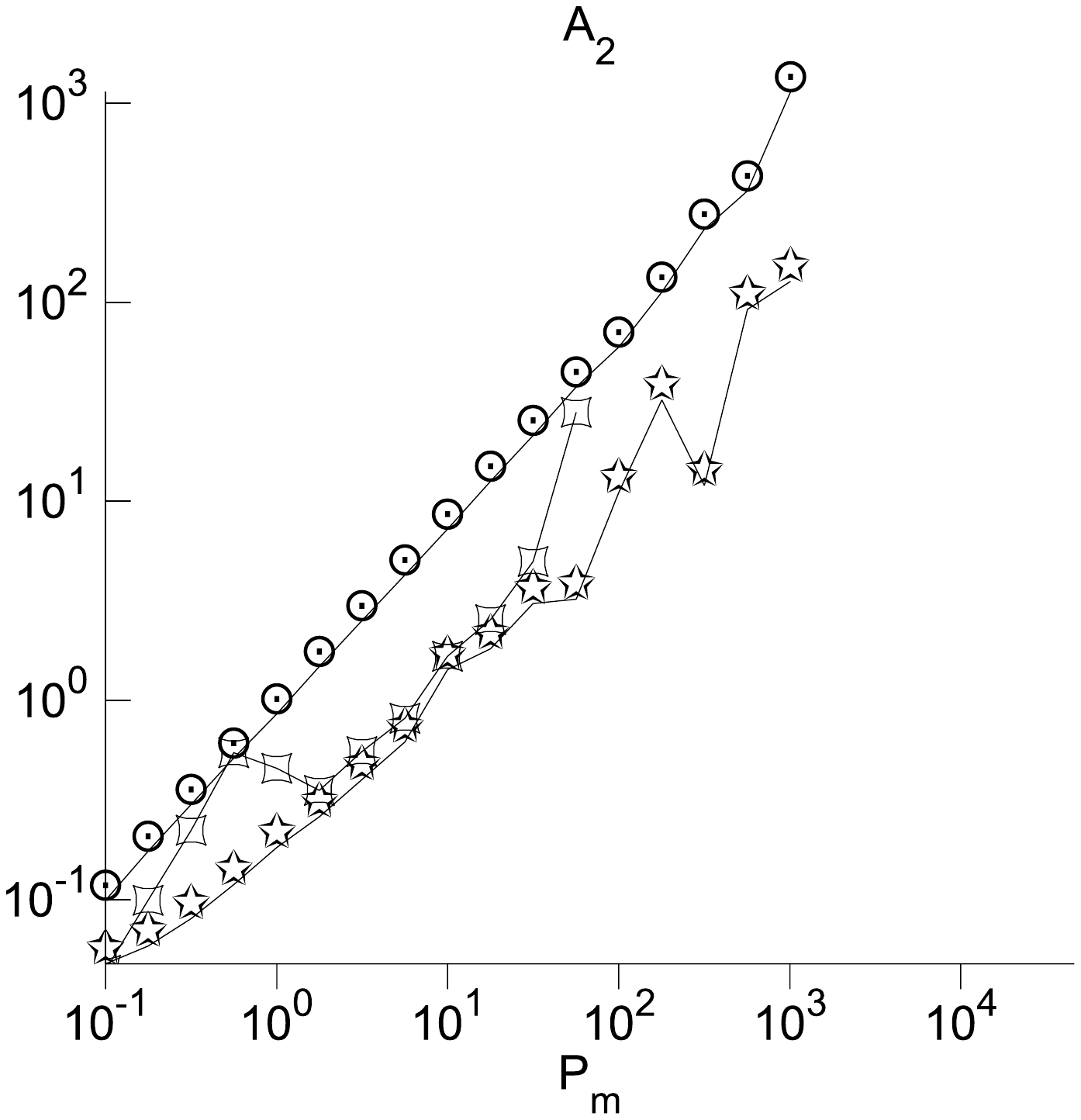}\includegraphics[width=.32\linewidth,angle=-90]{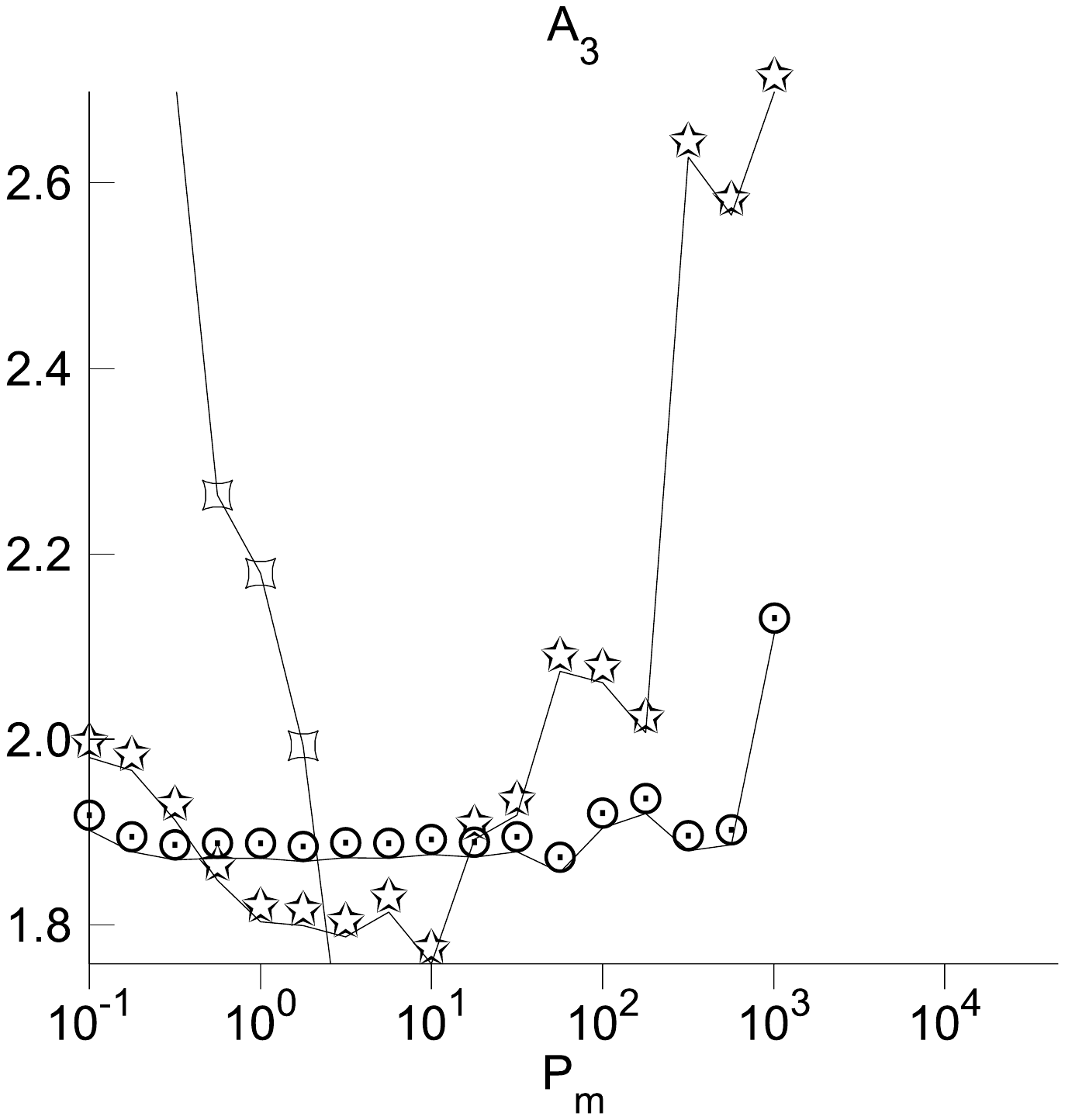} 
\par\end{centering}

\caption{\label{fig:A123} $A_{1,2,3}$ vs $P_{m}$. Top row: Case 1. The FOSA
solution is shown by circles, the exact solution by stars, and the
$\tau$ approximation by squares. The bottom row shows the same data
for Case 2.}

\end{figure}

To investigate the quenching of the $\alpha$-effect we need to examine
the dependence of $\mathcal{E}_{x}$ on $B_{x}$. We approximate this
with the following fitting functions, depending on three parameters
$A_{1},A_{2},A_{3}$: \begin{equation}
\frac{A_{1}B}{1+A_{2}B^{A_{3}}}\label{eq:fit}\end{equation}
 Examples of these fits for the different cases are shown on the fig\ref{fig:fits}.
The fit (\ref{eq:fit}) does not work well for high $P_{m}$ as the
mean electromotive force tends to zero and is highly fluctuating. However the limiting behaviour for strong magnetic fields is approximated
quite well. The deviation of the FOSA from the exact solution is clearly
seen for high conductivity and Case 2. In the same way we can say
that the $\tau$ approximation gives a better fit to the exact solution
for those parameter values. This is confirmed by the results shown
in Figure \ref{fig:A123}, where we show variations of $A_{1-3}$
with $P_{m}$.

Several features are quite well seen in Figure \ref{fig:A123}. First,
in Case 1 the $\tau$ approximation seems bad. Even the sign of effect
is opposite to that for the exact solution. On the other hand a significant difference between the exact solution
and FOSA is well seen for the high $P_{m}$. Second, {}``catastrophic
quenching'', when $A_{2}\sim R_{m}$, is found for the high-conducting
case. This phenomenon is more pronounced for FOSA than for the $\tau$
approximation and the full solution. Third, in Case 1 for FOSA the
power $A_{3}$ of quenching function is about $1.8$ in the whole range
while for Case 2 it is slightly higher - $2$. The quenching power
of the exact and FOSA solutions are close.

\begin{figure}
\begin{centering}
\includegraphics[height=.45\linewidth,angle=-90]{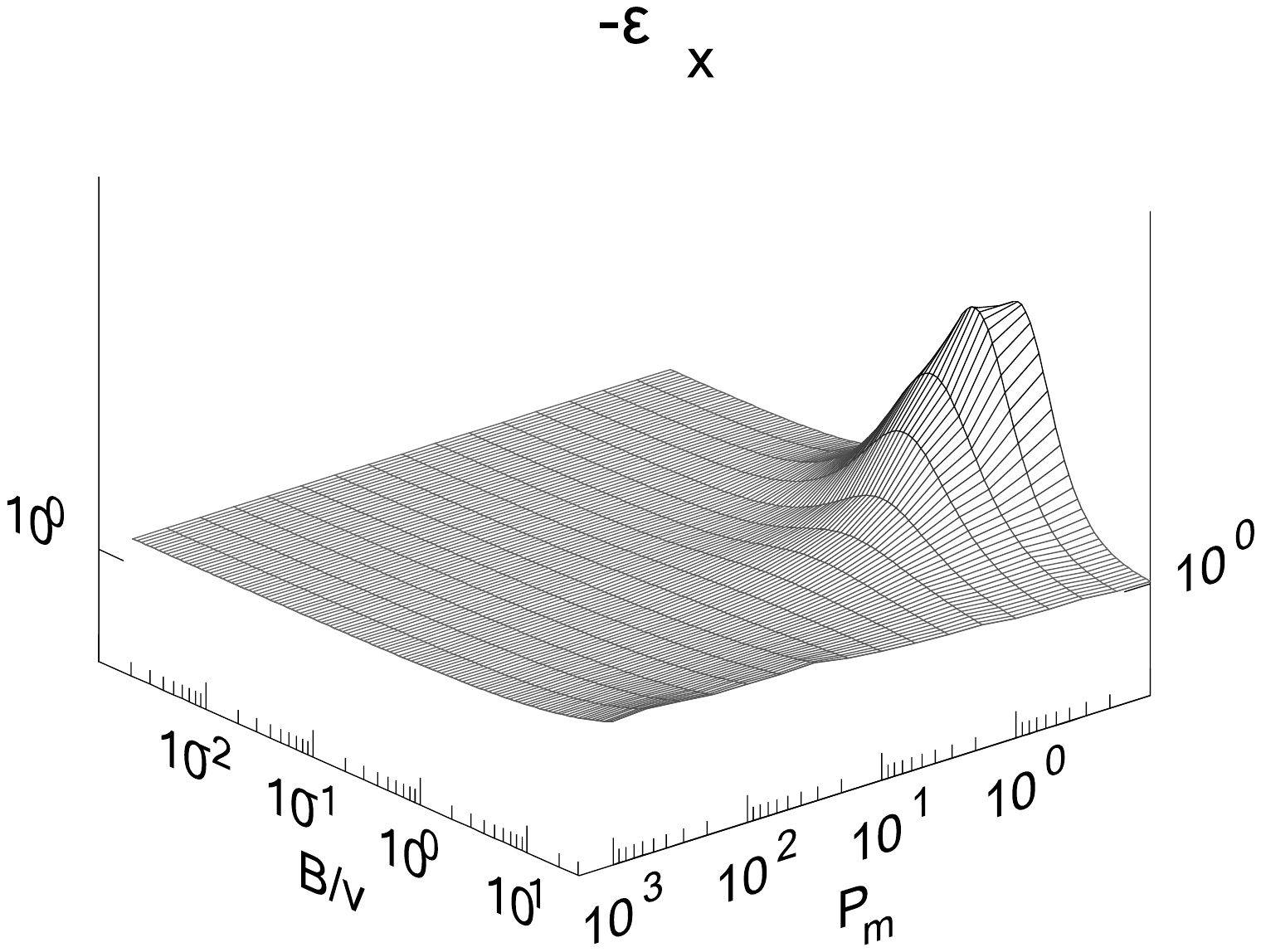}  \includegraphics[height=.45\linewidth,angle=-90]{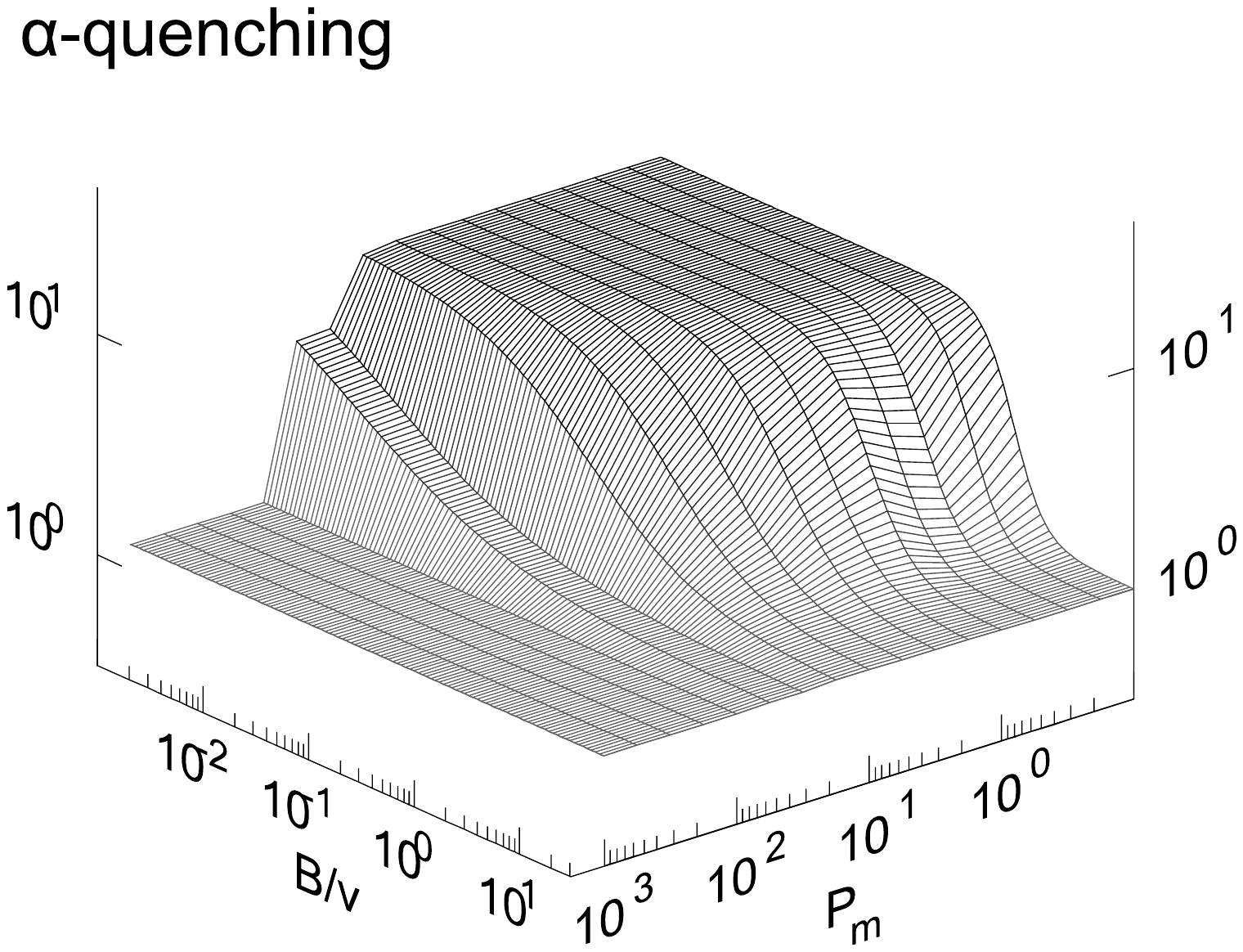} 
\end{centering}

\caption{\label{fig:3da}Plots of $-\mathcal{E}_x$ (left) and $\alpha$ (right) as functions of $B/v$ and $P_m$ for Case 2. Case 1 is similar (sf Figure\ref{fig:A123}).}

\end{figure}

Plots of the  amplitude of $\mathcal{E}_{x}$  and the alpha-quenching
as functions of $P_{m}$ and magnetic field strength $B/\nu$ are shown
in Figure \ref{fig:3da}. Again we see that {}``catastrophic'' quenching
occurs for high $P_{m}$.

\begin{figure}
\begin{centering}
\includegraphics[width=.5\linewidth,angle=-90]{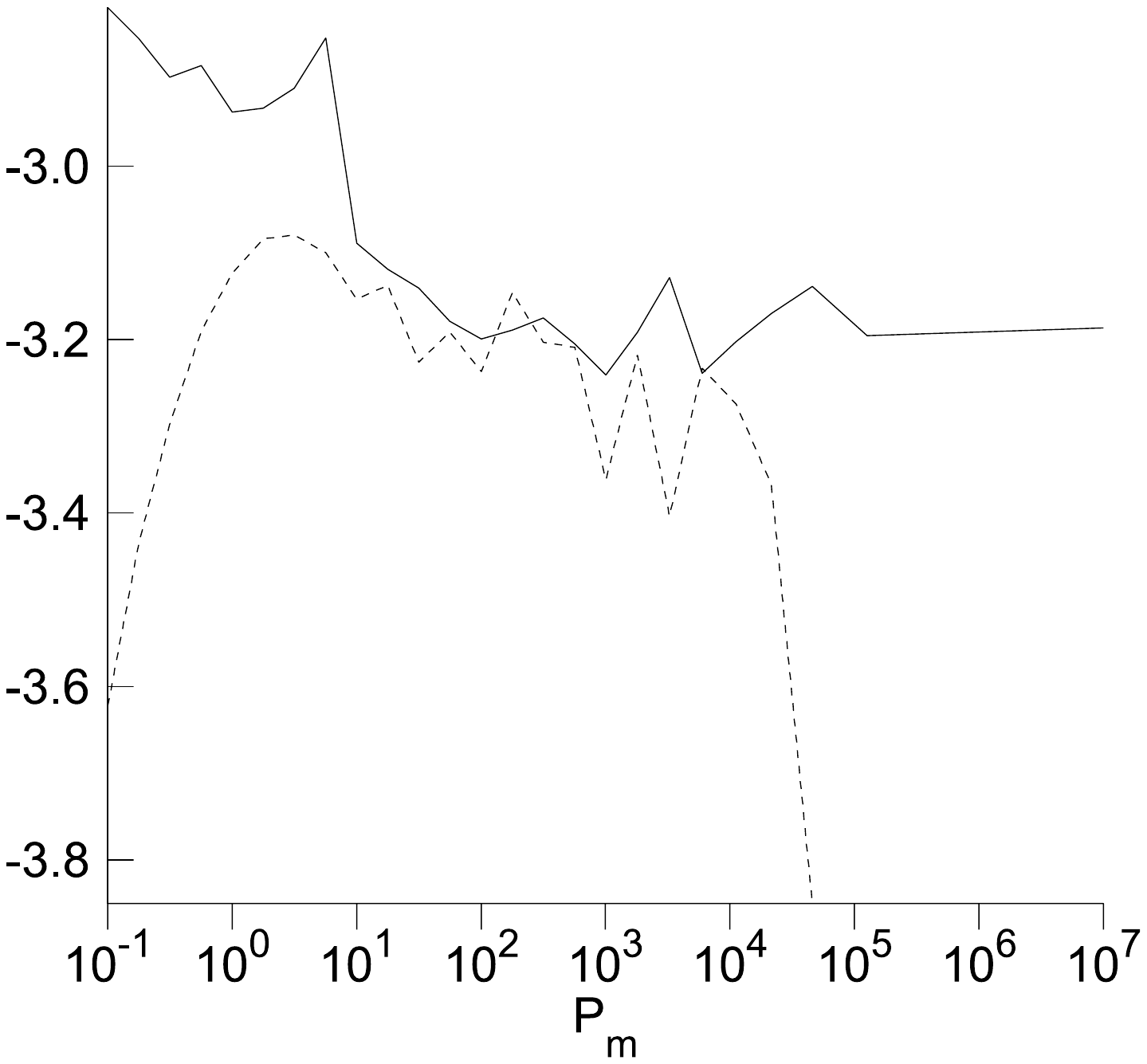}\includegraphics[width=.5\linewidth,angle=-90]{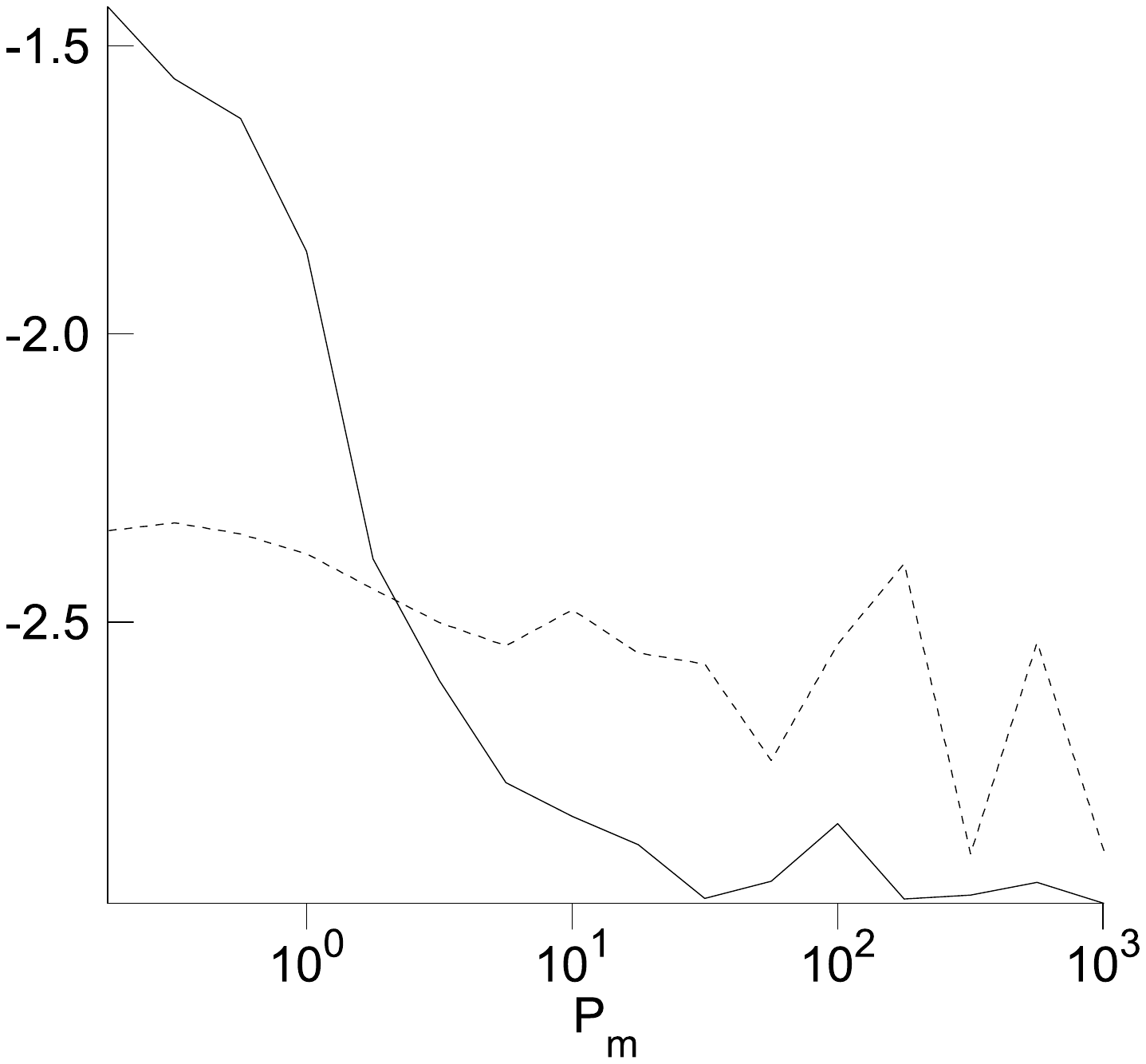} 
\par\end{centering}

\caption{\label{fig:hel}The alpha effect (dashed line) and residual helicity
$c\tau^{(h)}\left(h_{\mathcal{C}}-h_{\mathcal{K}}\right)$ (solid
line), for the steady forcing (left) and for the $\delta$-correlated
random force (right), as functions of $P_{m}$ and $B/\nu=0.001$. }

\end{figure}

A formula that is widely quoted and has been justified by use of the $\tau$
approximation is the simple relation between kinetic and current helicities
in turbulent flows and the $\alpha$ effect, $\alpha\sim\tau\left(h_{\mathcal{C}}-h_{\mathcal{K}}\right)$,
where $h_{\mathcal{C}}=\mu^{-1}\left\langle \mathbf{b\cdot\nabla\times b}\right\rangle $
and $h_{\mathcal{K}}=\left\langle \mathbf{u\cdot\nabla\times u}\right\rangle $\cite{bra-sub:04,krarad80,moff:78,kps:06}.
In Figure \ref{fig:hel} we show the $\alpha$ effect and residual
helicity $c\tau^{(h)}\left(h_{\mathcal{C}}-h_{\mathcal{K}}\right)$(with
$\tau^{(h)}$ as given in Figure \ref{fig:taus}) for two cases of
the random force driving the turbulence. The coefficient was approximately
chosen to match the maximum magnitude of the $\alpha$, we put $c=1/3$
both for Case 1 and for Case 2. Clearly, there is no unique relation
between $\alpha$ and residual helicity on the whole range of $P_{m}$.
Though there is correspondence in sign.

\begin{figure}
\begin{centering}
\includegraphics[width=.45\linewidth,angle=-90]{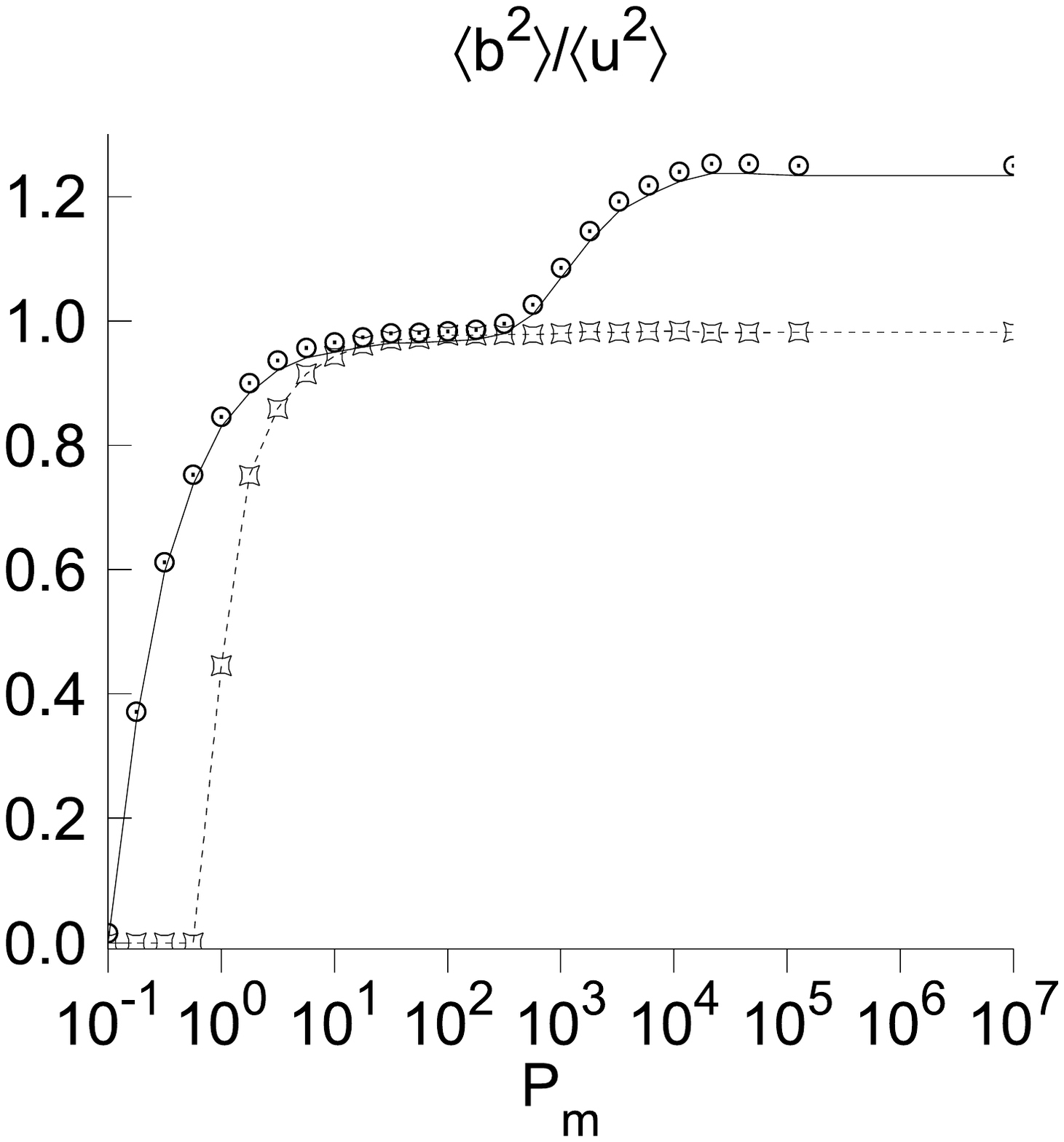}\ \  \includegraphics[width=.45\linewidth,angle=-90]{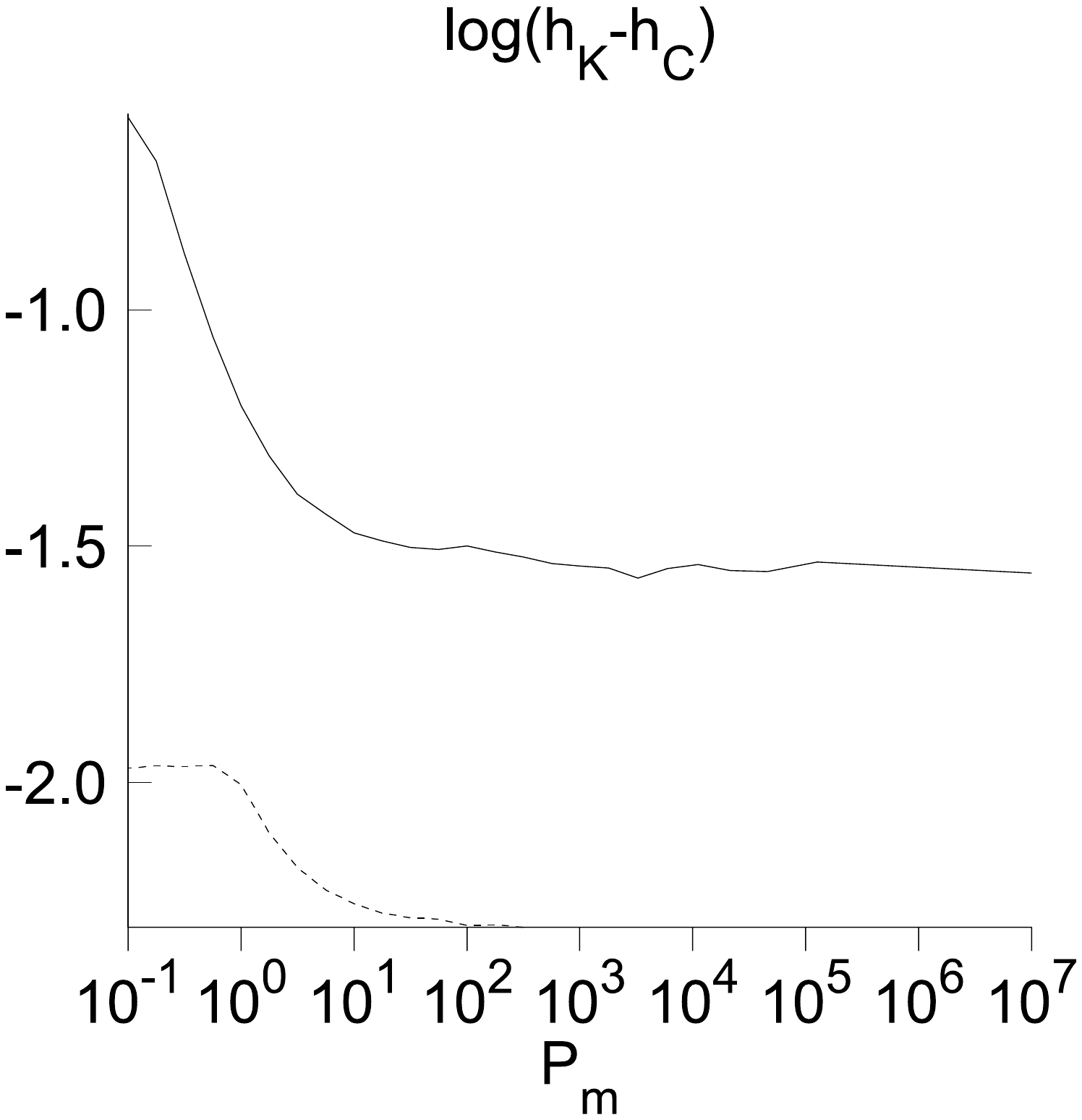}
\end{centering}

\caption{\label{f2-2}$\nu=0.01$,  
Left: ratio between magnetic and kinetic energy as a function of $P_m$. Squares are Case 1 and circles
are Case 2. Right: the residual helicity as a function of  $P_{m}$: Case 1,
dashed line, Case 2, solid line.}

\end{figure}

Next we consider some results for a somewhat higher Reynolds number with
$\nu=.01$. Again we present results for two cases. Case 1 is that
of zero correlation time: the force is updated at each timestep and
$Re\approx11$. In Case 2, which has finite correlation time, the
force was updated each 50-th time step and $Re\approx22.8$. The relation
between $P_{m}$ and the amplitude of the small-scale magnetic field
fluctuations for $\overline{B}=0$ is shown in Figure \ref{f2-2}.
For Case 1 there is a small-scale dynamo if $P_{m}\ge1$, while it
exists for $P_{m}>.1$ in Case 2 . Also in Case 2 we observe that
for the high enough $P_{m}$ the energy of magnetic fluctuations is slightly larger than its kinetic counterpart.

\begin{figure}
\begin{centering}
\includegraphics[width=.4\linewidth,angle=-90]{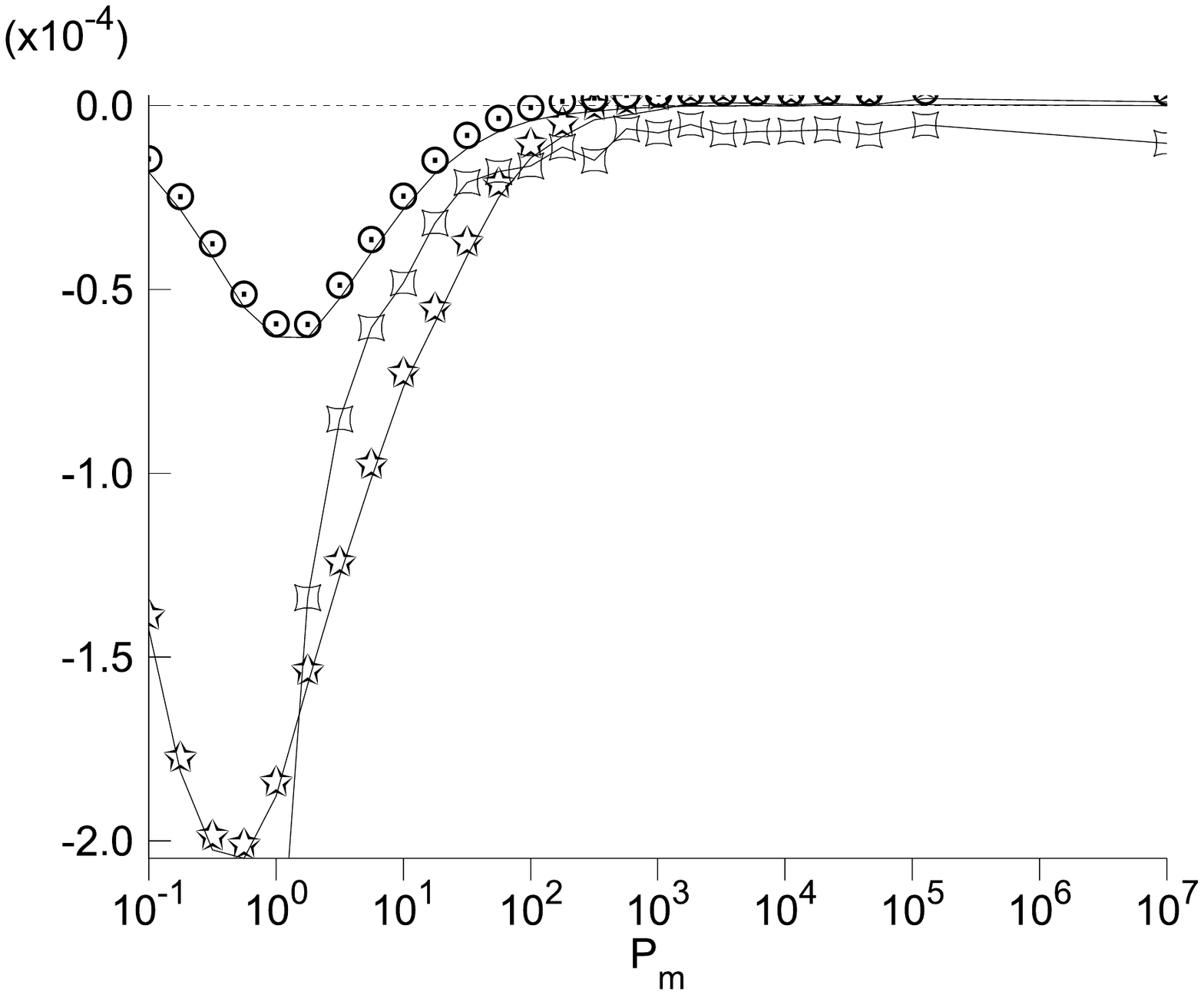}\includegraphics[width=.4\linewidth,angle=-90]{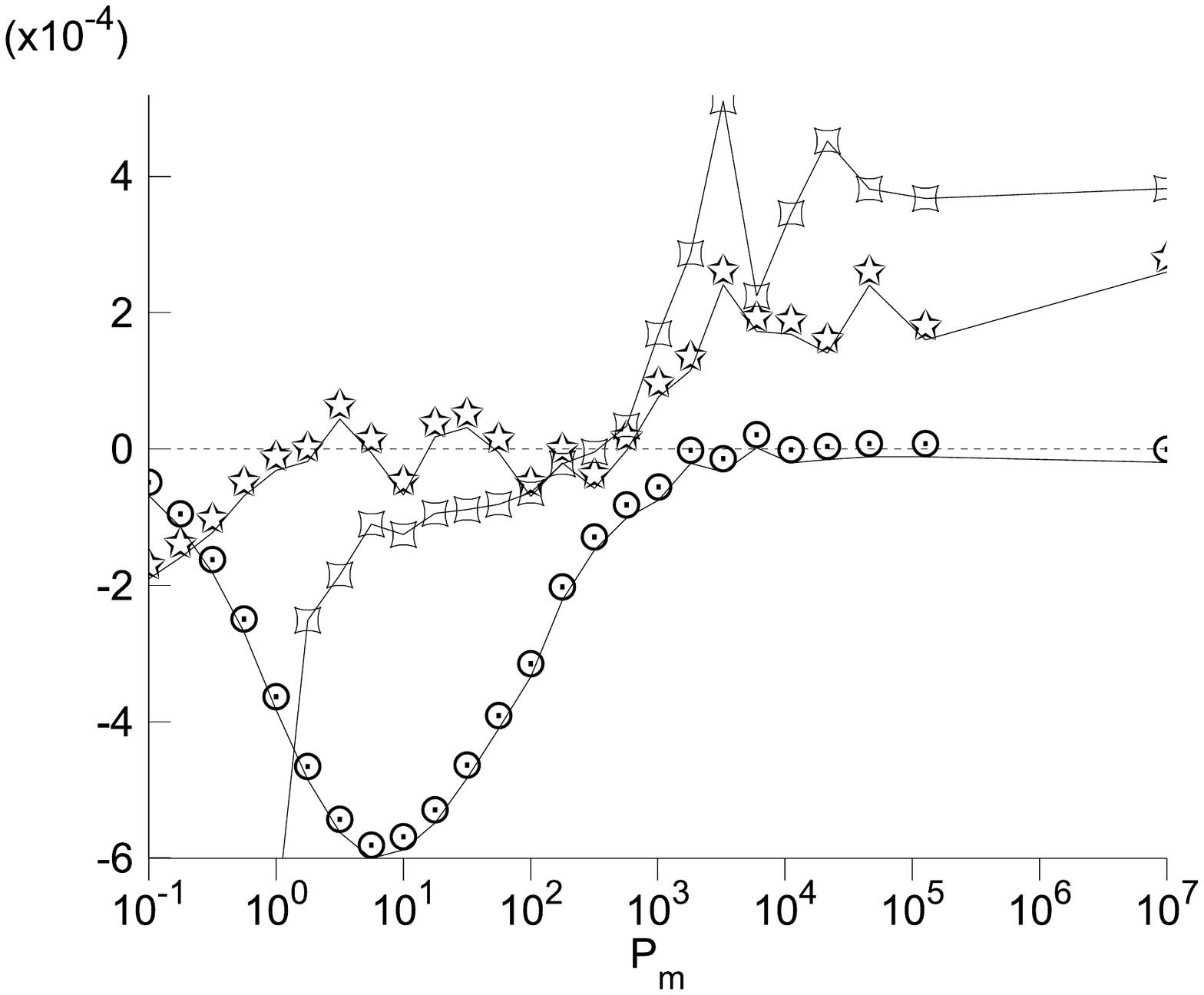} 
\par\end{centering}

\caption{\label{fig:hel2}$\nu=0.01$, the mean-electromotive force vs $P_{m}$
with mean field fixed to $B/\nu=0.01$, left is the the Case 1 right
- Case 2. We decrease the values of $\mathcal{E}$ obtained from FOSA
to factor 10 to make all the curves visible in one scale. Circles
show FOSA, stars - exact solution and squares - for the $\tau$-approximation. }

\end{figure}

This seems to be a main reason why the $\alpha$ effect changes 
sign as $P_{m}$ varies from low to high values.
Meanwhile the residual helicity ($\left(h_{\mathcal{C}}-h_{\mathcal{K}}\right)$)
does not. This is demonstrated on Figure\ref{fig:hel2} and helicity
is shown at right side on Figure\ref{f2-2}. The reversal of sign the
$\alpha$ effect for high $Rm$ was also found by \cite{CHT06}. Having
in mind that the energy of magnetic fluctuations dominates the kinetic
energy of the flow we could interpret this on the basis of results
of analytical calculations of the $\alpha$ effect for a rotating stratified
turbulence within $\tau$- approximation as those given in, e.g.,
\cite{rad-kle-rog,kle-rog:04c,pip07}. Suppose that the vector $\mathbf{U}$
characterizes the stratification scale and $\mathbf{\Omega}$ is a
global rotation velocity then for the case of slowly rotating media
penetrated with a weak large-scale magnetic field, within $\tau$-
approximation we obtain $\alpha\sim$$\left(\mathbf{\Omega\cdot U}\right)\tau_{c}^{2}\left({\displaystyle \frac{\left\langle b^{2}\right\rangle }{4\pi\rho}}-\left\langle u^{2}\right\rangle \right)$.
However in this theory the sign of expression in the brackets is intimately
related to the sign of the residual helicity which is not the case
for the  computational results presented above. This point needs further
clarifications in the multiscale model.

\section*{Discussion and conclusions}

One of the core issues of mean-field dynamo theory is the absence
of reliable method for evaluation of the kinetic coefficients which
describe the influence of turbulent dynamics on the evolution of the
large-scale field. This issue is related to the unsolved closure problem
in turbulence theories. Here we have attempted to construct a simple
nonlinear dynamical model that can be used for this purpose. The feasibility
of the model was demonstrated by numerical calculation of the nonlinear
$\alpha$ effect. Moreover, the model is helpful to check two basic
analytic {\it ans\"atze} of  mean-field magnetohydrodynamics - SOCA(FOSA)
and the $\tau$ approximation, at least for moderate values of $R_m$. Our results indicate that the $\tau$ approximation
may be useful in a dynamical regimes where the small-scale dynamo
is active. On the other hand, the results show  catastrophic quenching
of the $\alpha$ effect for high $P_{m}$. This is not found in analytic
computations either in \cite{kle-rog:04c} or in \cite{pip07}. Certainly
the applicability limits of this approximation need further clarification,
but we can say with some confidence that if the approximation schemes
fail for the present model they are unlikely to be very good for a
fully resolved calculation.

In the paper we present numerical calculations of the mean electromotive
force for two different temporal regimes of the random force driving
the turbulence. One case (Case 1) is essentially white-noise forcing
and the other (Case 2) is a coloured noise with a random force which
was updated each 50th time step (for our parameters this is about
two diffusion times of the system). We found that in the high conductivity
limit, the difference between SOCA and the full solution of the model
is quite significant. In particular, the full $\alpha$ effect is
more than 10 times smaller than that from SOCA. The difference in magnetic
quenching is not very large. For high $P_{m}$ the $\alpha$effect
is quenched $\alpha\sim\overline{B}^{-4}$ in the nonlinear model
though SOCA gives $\alpha\sim\overline{B}^{-3}$ which is consistent with previous findings by \cite{kit-rud:1993b,SSB07}.

The model is not competent to deal properly with turbulent diffusion
because there no energy transfer to different spatial scales. In fact,
it would be very useful to generalize the simple Fourier vector space
given at Figure \ref{fig1} to a more general one with several shells.
Then the effect of non-uniform magnetic fields and nonuniform flow
on the turbulence and the mean electromotive force can be investigated
in a similar way. One possible generalisation would be to consider a
decomposition of the fluctuating velocity of the form  
\[
\mathbf{u}\left(\mathbf{x}\right)=\sum_{j=1,3}\sum_{n=1,6}\mathbf{\hat{u}}^{(j,n)}e^{\imath(\mathbf{k}^{(j,n)}\cdot\mathbf{x})}+CC,\]
 where for the  superscripts $(j,n)$, $j=1,2,3$ is related to the number
of a vector shell and $n=1,\ldots 6$ is the number of a mode, and $CC$ denotes complex conjugate. Each shell is similar to that
of Figure \ref{fig1}. The modes of these shells interact in each  triplet since e.g. $\mathbf{k}^{(1,1)}+\mathbf{k}^{(2,1)}+\mathbf{k}^{(3,1)}=0$
and $\left|\mathbf{k}^{(1,1)}\right|\ne\left|\mathbf{k}^{(2,1)}\right|\ne\left|\mathbf{k}^{(3,1)}\right|$.
The dynamical system thus obtained obeys all conservation laws. It
should be suitable for the  evaluation of  $\alpha$ and other effects which
are important for the mean-field dynamo, e.g., turbulent diffusion,
or joint effect due to global rotation, nonuniform magnetic field
and nonuniform mean flow.

\subsection*{Acknowledgements} VVP thanks the Royal Society of London and Trinity College, Cambridge for financial support.
\bibliographystyle{plainnat}


\subsection*{Appendix A}

Nonlinear contributions in the induction eq.(\ref{eq:1}): \begin{eqnarray*}
\mathcal{M}_{i}^{\left(1\right)} & = & \imath k_{n}^{(1)}\left(\widetilde{\hat{b}}_{n}^{(2)}\hat{u}_{i}^{\left(3\right)}-\widetilde{\hat{b}}_{i}^{(2)}\hat{u}_{n}^{\left(3\right)}+\widetilde{\hat{u}}_{i}^{(2)}\hat{b}_{n}^{\left(3\right)}-\widetilde{\hat{u}}_{n}^{(2)}\hat{b}_{i}^{\left(3\right)}\right.\\
 &  & \left.+\widetilde{\hat{b}}_{n}^{(4)}\hat{u}_{i}^{\left(5\right)}-\widetilde{\hat{b}}_{i}^{(4)}\hat{u}_{n}^{\left(5\right)}+\widetilde{\hat{u}}_{i}^{(4)}\hat{b}_{n}^{\left(5\right)}-\widetilde{\hat{u}}_{n}^{(4)}\hat{b}_{i}^{\left(5\right)}\right),\\
\mathcal{M}_{i}^{\left(2\right)} & = & \imath k_{n}^{(2)}\left(\widetilde{\hat{b}}_{n}^{(1)}\hat{u}_{i}^{\left(3\right)}-\widetilde{\hat{b}}_{i}^{(1)}\hat{u}_{n}^{\left(3\right)}+\widetilde{\hat{u}}_{i}^{(1)}\hat{b}_{n}^{\left(3\right)}-\widetilde{\hat{u}}_{n}^{(1)}\hat{b}_{i}^{\left(3\right)}\right.\\
 &  & \left.+\widetilde{\hat{b}}_{n}^{(6)}\hat{u}_{i}^{\left(4\right)}-\widetilde{\hat{b}}_{i}^{(6)}\hat{u}_{n}^{\left(4\right)}+\widetilde{\hat{u}}_{i}^{(6)}\hat{b}_{n}^{\left(4\right)}-\widetilde{\hat{u}}_{n}^{(6)}\hat{b}_{i}^{\left(4\right)}\right),\\
\mathcal{M}_{i}^{\left(3\right)} & = & \imath k_{n}^{(3)}\left(\hat{b}_{n}^{(1)}\hat{u}_{i}^{\left(2\right)}-\hat{b}_{i}^{(1)}\hat{u}_{n}^{\left(2\right)}+\hat{u}_{i}^{(1)}\hat{b}_{n}^{\left(2\right)}-\hat{u}_{n}^{(1)}\hat{b}_{i}^{\left(2\right)}\right.\\
 &  & \left.+\widetilde{\hat{b}}_{n}^{(6)}\hat{u}_{i}^{\left(5\right)}-\widetilde{\hat{b}}_{i}^{(6)}\hat{u}_{n}^{\left(5\right)}+\widetilde{\hat{u}}_{i}^{(6)}\hat{b}_{n}^{\left(5\right)}-\widetilde{\hat{u}}_{n}^{(6)}\hat{b}_{i}^{\left(5\right)}\right),\\
\mathcal{M}_{i}^{\left(4\right)} & = & \imath k_{n}^{(4)}\left(\hat{b}_{n}^{(6)}\hat{u}_{i}^{\left(2\right)}-\hat{b}_{i}^{(6)}\hat{u}_{n}^{\left(2\right)}+\hat{u}_{i}^{(6)}\hat{b}_{n}^{\left(2\right)}-\hat{u}_{n}^{(6)}\hat{b}_{i}^{\left(2\right)}\right.\\
 &  & \left.+\widetilde{\hat{b}}_{n}^{(1)}\hat{u}_{i}^{\left(5\right)}-\widetilde{\hat{b}}_{i}^{(1)}\hat{u}_{n}^{\left(5\right)}+\widetilde{\hat{u}}_{i}^{(1)}\hat{b}_{n}^{\left(5\right)}-\widetilde{\hat{u}}_{n}^{(1)}\hat{b}_{i}^{\left(5\right)}\right),\\
\mathcal{M}_{i}^{\left(5\right)} & = & \imath k_{n}^{(5)}\left(\hat{b}_{n}^{(1)}\hat{u}_{i}^{\left(4\right)}-\hat{b}_{i}^{(1)}\hat{u}_{n}^{\left(4\right)}+\hat{u}_{i}^{(1)}\hat{b}_{n}^{\left(4\right)}-\hat{u}_{n}^{(1)}\hat{b}_{i}^{\left(4\right)}\right.\\
 &  & \left.+\hat{b}_{n}^{(3)}\hat{u}_{i}^{\left(6\right)}-\hat{b}_{i}^{(3)}\hat{u}_{n}^{\left(6\right)}+\hat{u}_{i}^{(3)}\hat{b}_{n}^{\left(6\right)}-\hat{u}_{n}^{(3)}\hat{b}_{i}^{\left(6\right)}\right),\\
\mathcal{M}_{i}^{\left(6\right)} & = & \imath k_{n}^{(6)}\left(\widetilde{\hat{b}}_{n}^{(2)}\hat{u}_{i}^{\left(4\right)}-\widetilde{\hat{b}}_{i}^{(2)}\hat{u}_{n}^{\left(4\right)}+\widetilde{\hat{u}}_{i}^{(2)}\hat{b}_{n}^{\left(4\right)}-\widetilde{\hat{u}}_{n}^{(2)}\hat{b}_{i}^{\left(4\right)}\right.\\
 &  & \left.+\widetilde{\hat{b}}_{n}^{(3)}\hat{u}_{i}^{\left(5\right)}-\widetilde{\hat{b}}_{i}^{(3)}\hat{u}_{n}^{\left(5\right)}+\widetilde{\hat{u}}_{i}^{(3)}\hat{b}_{n}^{\left(5\right)}-\widetilde{\hat{u}}_{n}^{(3)}\hat{b}_{i}^{\left(5\right)}\right),\end{eqnarray*}
 The nonlinear parts of momentum equation (\ref{eq:2}): \begin{eqnarray*}
\mathcal{N}_{i}^{\left(1\right)} & = & \imath k_{n}^{(1)}\left(\widetilde{\hat{b}}_{n}^{(2)}\hat{b}_{i}^{\left(3\right)}+\widetilde{\hat{b}}_{i}^{(2)}\hat{b}_{n}^{\left(3\right)}+\widetilde{\hat{b}}_{n}^{(4)}\hat{b}_{i}^{\left(5\right)}+\widetilde{\hat{b}}_{i}^{(4)}\hat{b}_{n}^{\left(5\right)}\right.\\
 &  & \left.-\widetilde{\hat{u}}_{n}^{(2)}\hat{u}_{i}^{\left(3\right)}-\widetilde{\hat{u}}_{i}^{(2)}\hat{u}_{n}^{\left(3\right)}-\widetilde{\hat{u}}_{n}^{(4)}\hat{u}_{i}^{\left(5\right)}-\widetilde{\hat{u}}_{i}^{(4)}\hat{u}_{n}^{\left(5\right)}\right),\\
\mathcal{N}_{i}^{\left(2\right)} & = & \imath k_{n}^{(2)}\left(\widetilde{\hat{b}}_{n}^{(1)}\hat{b}_{i}^{\left(3\right)}+\widetilde{\hat{b}}_{i}^{(1)}\hat{b}_{n}^{\left(3\right)}+\widetilde{\hat{b}}_{n}^{(6)}\hat{b}_{i}^{\left(4\right)}+\widetilde{\hat{b}}_{i}^{(6)}\hat{b}_{n}^{\left(4\right)}\right.\\
 &  & \left.-\widetilde{\hat{u}}_{n}^{(1)}\hat{u}_{i}^{\left(3\right)}-\widetilde{\hat{u}}_{i}^{(1)}\hat{u}_{n}^{\left(3\right)}-\widetilde{\hat{u}}_{n}^{(6)}\hat{u}_{i}^{\left(4\right)}-\widetilde{\hat{u}}_{i}^{(6)}\hat{u}_{n}^{\left(4\right)}\right),\\
\mathcal{N}_{i}^{\left(3\right)} & = & \imath k_{n}^{(3)}\left(\hat{b}_{n}^{(1)}\hat{b}_{i}^{\left(2\right)}+\hat{b}_{i}^{(1)}\hat{b}_{n}^{\left(2\right)}+\widetilde{\hat{b}}_{n}^{(6)}\hat{b}_{i}^{\left(5\right)}+\widetilde{\hat{b}}_{i}^{(6)}\hat{b}_{n}^{\left(5\right)}\right.\\
 &  & \left.-\hat{u}_{n}^{(1)}\hat{u}_{i}^{\left(2\right)}-\hat{u}_{i}^{(1)}\hat{u}_{n}^{\left(2\right)}-\widetilde{\hat{u}}_{n}^{(6)}\hat{u}_{i}^{\left(5\right)}-\widetilde{\hat{u}}_{i}^{(6)}\hat{u}_{n}^{\left(5\right)}\right),\\
\mathcal{N}_{i}^{\left(4\right)} & = & \imath k_{n}^{(4)}\left(\widetilde{\hat{b}}_{n}^{(1)}\hat{b}_{i}^{\left(5\right)}+\widetilde{\hat{b}}_{i}^{(1)}\hat{b}_{n}^{\left(5\right)}+\hat{b}_{n}^{(6)}\hat{b}_{i}^{\left(2\right)}+\hat{b}_{i}^{(6)}\hat{b}_{n}^{\left(2\right)}\right.\\
 &  & \left.-\widetilde{\hat{u}}_{n}^{(1)}\hat{u}_{i}^{\left(5\right)}-\widetilde{\hat{u}}_{i}^{(1)}\hat{u}_{n}^{\left(5\right)}-\hat{u}_{n}^{(6)}\hat{u}_{i}^{\left(2\right)}-\hat{u}_{i}^{(6)}\hat{u}_{n}^{\left(2\right)}\right),\\
\mathcal{N}_{i}^{\left(5\right)} & = & \imath k_{n}^{(5)}\left(\hat{b}_{n}^{(1)}\hat{b}_{i}^{\left(4\right)}+\hat{b}_{i}^{(1)}\hat{b}_{n}^{\left(4\right)}+\hat{b}_{n}^{(6)}\hat{b}_{i}^{\left(3\right)}+\hat{b}_{i}^{(6)}\hat{b}_{n}^{\left(3\right)}\right.\\
 &  & \left.-\hat{u}_{n}^{(1)}\hat{u}_{i}^{\left(4\right)}-\hat{u}_{i}^{(1)}\hat{u}_{n}^{\left(4\right)}-\hat{u}_{n}^{(6)}\hat{u}_{i}^{\left(3\right)}-\hat{u}_{i}^{(6)}\hat{u}_{n}^{\left(3\right)}\right),\\
\mathcal{N}_{i}^{\left(6\right)} & = & \imath k_{n}^{(6)}\left(\widetilde{\hat{b}}_{n}^{(3)}\hat{b}_{i}^{\left(5\right)}+\widetilde{\hat{b}}_{i}^{(3)}\hat{b}_{n}^{\left(5\right)}+\widetilde{\hat{b}}_{n}^{(2)}\hat{b}_{i}^{\left(4\right)}+\widetilde{\hat{b}}_{i}^{(2)}\hat{b}_{n}^{\left(4\right)}\right.\\
 &  & \left.-\widetilde{\hat{u}}_{n}^{(2)}\hat{u}_{i}^{\left(4\right)}-\widetilde{\hat{u}}_{i}^{(2)}\hat{u}_{n}^{\left(4\right)}-\widetilde{\hat{u}}_{n}^{(3)}\hat{u}_{i}^{\left(5\right)}-\widetilde{\hat{u}}_{i}^{(3)}\hat{u}_{n}^{\left(5\right)}\right),\end{eqnarray*}
\label{lastpage}
\end{document}